\documentclass[onecolumn,notitlepage]{article}
\usepackage{graphicx}
\usepackage{amsmath}
\usepackage{amsfonts}
\usepackage{amssymb}
\newtheorem{theorem}{Theorem}

\newtheorem{corollary}[theorem]{Corollary}

\begin{document}

\title{General Mechanism for a Positive Temperature Entropy Crisis in Stationary
Metastable States: Thermodynamic Necessity and Confirmation by Exact calculations}
\author{P. D. Gujrati\\Department of Physics, Department of Polymer Science, \\The University of Akron, Akron, OH 44325, USA}
\date{\today}
\maketitle
\begin{abstract}
We introduce the concept of stationary metastable states (SMS's), and give a
prescription to study it using a restricted partition function formalism. This
requires introducing a continuous entropy function $S(E)$ even for a finite
system, a standard practice in the literature though never clearly stated, so
that it can be differentiated. The formalism ensures that SMS free energy
exists all the way to $T=0$, and remains stable. We introduce the concept of
the reality condition, according to which the entropy $S(T)$\ of a set of
coupled degrees of freedom must be non-negative.The entropy crisis, which does
not affect stability, is identified as the violation of the reality condition.
We identify and validate rigorously, using general thermodynamic arguments,
the following general thermodynamic mechanism behind the entropy crisis in
SMS. The free energy $F_{\text{dis}}(T)$\ of any SMS must be equal to the
$T=0$ crystal free energy $E_{0}$ at two different temperatures $T=0,$\ and
$T=T_{\text{eq}}>0$. Thus, the stability requires $F_{\text{dis}}(T)$\ to
possess a maximum at an intermediate but a strictly positive temperature
$T_{\text{K}},$ where the energy is $E=E_{\text{K}}.$ The SMS branch below
$T_{\text{K}}$ gives the entropy crisis and must be replaced by hand by an
ideal glass free energy of constant energy $E_{\text{K}},$ and vanishing
entropy. Hence, $T_{\text{K}}>0$ represents the Kauzmann temperature. The
ideal glass energy $E_{\text{K}}$\ is higher than the crystal energy $E_{0}$
at absolute zero, which is in agreement with the experimenatal fact that the
extrapolated energy of a real glass at $T=0$ is higher than its $T=0$ crystal
energy. We confirm the general predictions by two exact calculations, one of
which is not mean-field. The calculations clearly show that the notion of SMS
is not only not vaccuous, but also not a consequence of a mean-field analysis.
They also show that certain folklore cannot be substantiated.
\end{abstract}

\section{Introduction}

Glass transition is a ubiquitous phenomenon$\cite{Kauzmann,Goldstein},$
believed to occur in a supercooled liquid (SCL), which is one of the
metastable states in Nature, and has been investigated for at least over eight
decades since the earliest works of Nernst$\cite{Nernst}$ and
Simon$\cite{Simon}$. Despite this, a complete understanding of the transition
itself and related issues is still far from complete, although major progress
has been made recently\cite{Stillinger,Sciortino,GujCorsi,GujRC}. Theoretical
and experimental investigations invariably require applying (time-independent)
thermodynamics to SCL to extract quantities like the entropy. This presumes,
as we do here, that there exists a \textit{stationary limit} of the metastable
SCL state (SMS) under infinitely slow cooling in which the crystal
(CR)\ is\textit{\ }forbidden from nucleating$\cite{Debenedetti}.$ We further
assume that the SCL free energy can be defined, at least mathematically (see
below for details), all the way down to absolute zero, which may not always be
possible$.$ It was discovered recently that under some conditions, SCL free
energy can terminate in a spinodal at a non-zero temperature as we lower the
temperature$\cite{GujRC}.$

Experimentally, the SCL configurational entropy exhibits a rapid drop near a
temperature $\simeq$ two-thirds of the melting temperature $T_{\text{M}}
\cite{Kauzmann,Goldstein}$. It is found that the smooth low-temperature
\textit{extrapolation} of the measured excess entropy $S_{\text{ex}}(T)\equiv
S_{\text{SCL}}(T)-S_{\text{CR}}(T)$ becomes negative$\cite{Kauzmann}$ below a
non-zero temperature. Since it is hard to imagine CR, being more ordered,
having a larger entropy than SCL, Kauzmann suggested that something like a
glass transition must intervene to avoid this entropy crisis, known commonly
as the \emph{Kauzmann paradox} ($S_{\text{ex}}(T)<0)\cite{Kauzmann},$ at some
positive temperature. There are several computational$\cite{Sciortino}$ and
theoretical$\cite{GujCorsi,GujRC,GibbsDi,Derrida}$ results clearly
demonstrating the existence of some kind of entropy crisis.

It should be stressed that there is no thermodynamic requirement for
$S_{\text{ex}}(T)$ to be non-negative. There are physical systems like
He$^{4}$ in which $S_{\text{ex}}(T)$ can become negative at low temperatures.
A recent exact calculation on a classical system$\cite{GujCorsi}$ clearly
demonstrates a negative $S_{\text{ex}}(T)$ at low temperatures$.$ If there is
any hope of finding a thermodynamic basis for the glass transition, we must
look for a condition for the glass transition that is dictated by
thermodynamics. Thus, in the following, we replace $S_{\text{ex}}(T)$ by the
entropy $S(T),$ such as the configurational entropy, which represents the
natural log of the \emph{number} of microstates $W(E),$ where $E$ is the
average energy at that temperature. Consequently, a state with negative
entropy is impossible to observe in Nature. The violation of the \emph{reality
condition} $S(T)\geq0$ signifies a genuine or absolute \emph{entropy crisis}
in Nature. We will interpret the entropy crisis in this work to signify the
reality condition violation $S(T)<0,$ and denote the temperature by
$T_{\text{K}},$ the Kauzmann temperature, when the violation begins as the
temperature is reduced.

It is widely recognized that $T_{\text{K}}$ is a theoretical point and not
accessible by experiment. However, its accessibility in itself is not
important if its usage helps us understand or explain glassy behavior. It is
well known that absolute zero is inaccessible; yet the study of a statistical
model at $T=0$ is a time-honored first step to study the physics of the model
at higher temperatures. The concept of a Kauzmann temperature enables us to
explain many glassy behavior. For example, the existence of the entropy crisis
at a non-zero temperature is conceptually necessary for the observed
Vogel-Fulcher behavior in fragile systems. Its importance cannot be denied
just because it is experimentally inaccessible. A majority of the experts in
the field continues to find the Kauzmann temperature to be an extremely useful
concept. The most important consequence of a positive $T_{\text{K}}$ is that
there is a lower limit to supercooling before an experimental glass transition
must intervene, and this lower limit is not absolute zero. This is an
extremely useful information for experimentalists who are interested in
investigating glassy behavior.

The following two experimental observations play a very important role in our
understanding of the glass transition:

(Expt1) The heat capacity of the glassy state can be substantially different
from that of the corresponding crystal at the same temperature\cite{GujGold}.

(Expt2) The energy of the glassy state ($E_{\text{K}})$ after extrapolation to
absolute zero is higher than that of the CR energy $E_{\text{0}}%
\cite{Kauzmann,Goldstein}$.\ \ \ \ \ \ \ \ \ 

Based on these observations, we have recently proposed$\cite{GujCorsi,GujRC}$
a mechanism behind the entropy crisis and the ideal glass transition in the
SMS of a system of long polymers. The proposed mechanism is as follows.

\textbf{Entropy Crisis} \textbf{Mechanism\ }\emph{Since }$E_{\text{K}}>E_{0}%
,$\emph{\ the SMS free energy rises rapidly below the melting temperature
}$T_{\text{M}}$\emph{\ as the system is cooled and crosses the }%
($T=0)-$\emph{CR free energy }$E_{0}$\emph{\ at some positive temperature
}0\emph{%
$<$%
}$T_{\text{eq}}<T_{\text{M}}.$\emph{\ The SMS free energy again equals }$E_{0}
$\emph{\ at }$T=0$\emph{. Therefore, it must go through a maximum at }0\emph{%
$<$%
}$T_{\text{K}}<T_{\text{eq}}$\emph{, at which the entropy of the system
vanishes. Since negative entropy below }$T_{\text{K}}$\emph{\ is not possible
for physical states, the SMS below }$T_{\text{K}}$\emph{\ is replaced by an
ideal glass with a constant free energy\ }$E_{\text{K}}$\emph{, and zero
entropy and heat capacity.}

Since the above two experimental observations are the main ingredients for the
proposed mechanism, it appears likely that the mechanism is common to all
glass forming systems with metastability, and not just restricted to long
polymers. If true, this will suggest that the above mechanism is a generic
root cause of the entropy crisis at a positive temperature in all systems in
which glasses are formed by cooling their metastable states.

However, the situation has become very confusing, as there have appeared
several recent arguments$\cite{Stillinger1,Kivelson,Johari,Wunderlich}$
against the existence of the entropy crisis. These arguments contradict many
exact calculations$\cite{GujCorsi,GujRC,GibbsDi,Tonks,Derrida}$ and
simulations$\cite{Sciortino}$ that clearly establish a positive-temperature
entropy crisis. We should also mention classical real gases, like the van der
Waals gas, or the classical ideal gas that are known to give entropy crisis at
positive temperatures\cite{Note2}. It is, therefore, extremely important to
clarify the issue, which we do by proving the general validity of the above
mechanism behind all glass formation in metastable states. Two important
aspects in the mechanism require general justification\cite{GujProof}:

(M1) The SMS free energy, obtained by mathematical continuation of the
disordered equilibrium state below the melting temperature,\ and the CR free
energy are the same at absolute zero.

(M2) The temperature $T_{\text{K}}$ when the entropy crisis occurs first is non-zero.

Together, they support the existence of an \emph{entropy crisis} at a positive
temperature. The importance of a mathematical proof is that it settles the
issue once for all. However, another very important consequence of the proof
is, as we will see here, that if a phase in some calculation gives rise to an
entropy crisis, it means that there must be another (ordered) phase without an
entropy crisis. The proof uses general but rigorous thermodynamic and
statistical mechanical arguments, valid for classical or quantum systems. We
verify our conclusions by two exact calculations, one of which is not a
mean-field calculation.\ 

As the issue has remained unresolved for so long most certainly implies that
its resolution is not trivial.\emph{\ }Thus, it should come as no surprise to
the reader that our arguments are somewhat involved, can be divided into
several clear parts, and require patience to go through. However, we believe
that they are certainly not beyond the reach of a majority of the workers
studying glasses.\ The proof of the validity of the above mechanism follows
from the various theorems that we prove here. To guide the reader, we
summarize the strategy for the eventual demonstration.

1. We first prove that for $E_{\text{D}}>E_{0},$ where $E_{\text{D}}$ is the
lowest energy for which SMS exists $[S(E)\geq0]$, the SMS free energy must
become equal to the ($T=0)$- CR free energy $E_{0}$ at a positive temperature
$T_{\text{eq}}.$ Neither the SMS entropy nor its inverse slope at
$E_{\text{D}}$ need vanish for this to hold.

2. We then prove that if the energy of the SMS state is higher than $E_{0},$
then its temperature must be positive. From this it follows that SMS cannot
reach $T=0$ with an energy $E=E_{\text{D}}>E_{0}$.

3. We then prove that SMS free energy at $T=0$ again equals $E_{0}$\ at $T=0$.
Thus, the SMS free energy must have a maximum somewhere in the range
$0<T<T_{\text{eq}}.$ The entropy at the maximum is zero, identifying the
maximum with the Kauzmann point and its temperature with $T_{\text{K}}$. This
also identifies the lowest energy $E_{\text{D}}$ with $E_{\text{K}}$, the
energy at $T_{\text{K}}.$ This energy is greater than $E_{0}$ because of the
non-negative SMS heat capacity.

The existence of the maximum in the SMS free energy thus finally proves the
existence of a Kauzmann point.

\subsection{Fundamental Postulate}

Our general proof assumes the \emph{existence} of SMS's, so that
thermodynamics can be applied. The need for the assumption is easy to
understand. At present, our understanding of whether equilibrium states can be
demonstrated to exist mathematically even in simple models is too limited. We
should recall that the existence of equilibrium states is taken for granted as
a \emph{postulate} in statistical mechanics and thermodynamics, where it is
well known that it is extremely hard to prove their existence. We quote
Huang$\cite{Huang}$:\ ``Statistical mechanics, however, does not describe how
a system approaches equilibrium, nor does it determine whether a system can
ever be found to be in equilibrium. It merely states what the equilibrium
situation is for a given system.'' Ruelle$\cite{Ruelle}$ notes that
equilibrium states are defined \emph{operationally} by assuming that the state
of an isolated system tends to an equilibrium state as time tends to
+$\infty.$ Whether a real system actually approaches this state cannot be
answered. We make a similar assumption about the existence of SMS within the
\emph{restricted} ensemble that will be introduced below. Any analysis of
thermal data on glasses and supercooled liquids requires using
time-independent thermodynamics, as discussed above. Thus,
\emph{hypothesizing} the existence of SMS is perfectly justified by the
practice in the field. The existence of supercooled liquids, and glasses in
systems with short- or long-range interactions or structural glasses, and
their stability over a long periods of time is undeniable. These stable states
are a manifestation of SMS in these systems.

Without such an assumption, we cannot justify using conventional
(time-independent) thermodynamics to analyze SCL data. The two exact
calculations in this work show that the hypothesis is not vacuous. The
behavior in real systems is, of course, oblivious to the state of our
knowledge and is not controlled by the fact that we can only demonstrate SMS
in a few exact calculations.

\subsection{Reality Condition and Entropy Crisis}

As said above, the entropy $S(T),$ such as the configurational entropy, that
we consider in this work represents the natural log of the number of
microstates $W(E)$ where $E$ is the average energy at that temperature. For
the microstates to exist in Nature, it is evident that $W(E)$ must satisfy the
\emph{reality condition}$\ W(E)\geq1$ [$S(T)\geq0$]. However, a state with
\emph{negative entropy} can emerge in theoretical calculations or
extrapolations. If it happens that the calculations or extrapolations result
in a negative $S(T)$ below some positive temperature $T_{\text{K}},$ this
should be interpreted as the absence of real microstates available to the
system at those temperatures and the system cannot be found in those
microstates in Nature. It is this violation of the reality condition
$S(T)\geq0$ that signifies the genuine entropy crisis in Nature.

It should be noted that our criterion for the entropy crisis is much more
stringent than the original Kauzmann requirement $S_{\text{ex}}(T)<0,$ since
it is possible to have $S(T)\geq0$ even though $S_{\text{ex}}(T)<0.$ Thus, our
$T_{\text{K}}$ is lower than the corresponding temperature where
$S_{\text{ex}}(T)=0.$

There are two $\emph{independent}$ aspects of thermodynamics and statistical
mechanics. The first one is the requirement of \emph{stability} according to
which thermodynamic quantities like the heat capacity, the compressibility,
etc. must never be negative. The other aspect, independent of the stability
criteria, is the \emph{reality} condition that ensures that such states occur
in Nature\cite{Note3}. The reality issue is central in our approach and is
discussed further below. We will see below that the mathematical extension of
the free energy of the disordered liquid phase below the melting temperature
$T_{\text{M}}$ will always satisfy the stability criteria everywhere ($T\geq
0$), but the reality condition is satisfied only above some positive
temperature $T_{\text{K}}.$

\subsection{Common Folklore}

Majority of the calculations displaying stationary metastable states are in
spin models, and are carried out at the level of the random-mixing
approximation (RMA); the latter is valid in the limit of infinite coordination
number $q$ and vanishing interaction strength $J,$ keeping $qJ$ is fixed and
finite. This approximation is equivalent to solving the models in an
\textit{infinite-dimensional }space. This has given rise to the common
folklore that SMS's occur only in an infinite-dimensional\textit{\ }space
whose coordination number is also infinite. This is incorrect as we will
clearly demonstrate here by an explicit calculation. The calculation is a
\emph{non-mean-field calculation }carried out exactly in a 1-dimensional model
and captures SMS.\emph{\ }The calculation is presented to overcome this
folklore. The model also shows the entropy crisis in SMS. Another alternative
way to visualize the RMA is to think of long-range interactions in a
finite-dimensional space. Thus, another folklore is that SMS does not exist
for short range models. Even frustration is considered in the folklore to be
necessary for the glassy behavior. The second example in the work is that of
an Ising model with \emph{short-range interactions and no frustration. }This
example establishes not only the existence of SMS in short range models but
also that frustration is not necessary for the existence. The model is solved
exactly on a Husimi cactus; hence, its thermodynamics is proper.

It is true that the cactus can only be embedded in an infinite dimensional
space, but locally it resembles a square lattice which has a finite
coordination number. The important point to note is that the interactions are
short-ranged. The dimensionality is relevant only if we are interested in
critical exponents, which we are not in this work. The \emph{finite
coordination number} of the cactus makes our calculation very different from
RMA. Our general proofs are certainly not based on RMA ideas. From all the
experience we have accumulated, the cactus provides a much better description
of the square lattice model than the conventional mean-field (RMA), as we have
shown elsewhere.$\cite{GujPRL}$\ \ \ \ \ \ \ \ \ \ \ \ \ \ \ \ \ \ \ \ \ \ \ \ \ \ \ \ \ \ \ \ \ \ \ \ \ \ \ \ \ \ \ \ \ \ \ \ \ \ \ \ \ \ \ \ \ \ \ \ \ \ \ \ \ \ \ \ \ \ \ \ \ \ \ \ \ \ \ \ \ \ \ \ \ \ \ \ \ \ \ \ \ \ \ \ \ \ \ \ \ \ \ \ \ \ \ \ \ \ \ \ \ \ \ \ \ \ \ \ \ \ \ \ \ \ \ \ \ \ \ \ \ \ \ \ \ \ \ \ \ \ \ \ \ \ \ \ \ \ \ \ \ \ \ \ \ \ \ \ \ \ \ \ \ \ \ \ \ \ \ \ \ \ \ \ \ \ \ \ \ \ \ \ \ \ \ \ \ \ \ \ \ \ \ \ \ \ \ \ \ \ \ \ \ \ \ \ \ \ \ \ \ \ \ \ \ \ \ \ \ \ \ \ \ \ \ \ \ \ \ \ \ \ \ \ \ \ \ \ \ \ \ \ \ \ \ \ \ \ \ \ \ \ \ \ \ \ \ \ \ \ \ \ \ \ \ \ \ \ \ \ \ \ \ \ \ \ \ \ \ \ \ \ \ \ \ \ \ \ \ \ \ \ \ \ \ \ \ \ \ \ \ \ \ \ \ \ \ \ \ \ \ \ \ \ \ \ \ \ \ \ \ \ \ \ \ \ \ \ \ \ \ \ \ \ \ \ \ \ \ \ \ \ \ \ \ \ \ \ \ \ \ \ \ \ \ \ \ \ \ \ \ \ \ \ \ \ \ \ \ \ \ \ \ \ \ \ \ \ \ \ \ \ \ \ \ \ \ \ \ \ \ \ \ \ \ \ \ \ \ \ \ \ \ \ \ \ \ \ \ \ \ \ \ \ \ \ \ \ \ \ \ \ \ \ \ \ \ \ \ \ \ \ \ \ \ \ \ \ \ \ \ \ \ \ \ \ \ \ \ \ \ \ \ \ \ \ \ \ \ \ \ \ \ \ \ \ \ \ \ \ \ \ \qquad

\section{Formulation}

\subsection{Canonical Partition Function}

The stationary nature of the SMS allows us to investigate it using the
partition function (PF) formalism. We consider a system composed of $N$
particles confined in a given volume $V$ and at a given temperature $T$. The
thermodynamic limit is obtained by taking the simultaneous limits $N$
$\rightarrow\infty$, and $V$ $\rightarrow\infty,$ such that the volume per
particle $v\equiv V/N$ is kept fixed and finite. The internal degrees of
freedom of the system contain the configurational (i.e., positional) degrees
of freedom and all other degrees of freedom like the translational degrees of
freedom \emph{coupled} to them\cite{Note1}. The canonical PF determined by
these degrees of freedom is given by
\begin{equation}
Z_{N}(T)\equiv\widehat{\text{Tr}}\;\exp(-\beta E), \label{dofZ}%
\end{equation}
where the trace operation $\widehat{\text{Tr}}\ $is over the coupled degrees
of freedom\ and $\beta\equiv1/T,$ $T$ being the system temperature in the
units of the Boltzmann constant $k_{\text{B}}.$ In the following, we will
usually suppress the index $N$ on the PF, unless necessary. The energy of the
system $E$ is determined by the coupled degrees of freedom. The PF in
(\ref{dofZ}) is \emph{irreducible} in that it cannot be written as a product
of several non-trivial PF's corresponding to independent sets of degrees of
freedom; see \cite{Note1}. Some coupling between different degrees of freedom,
no matter how weak, is required to achieve equilibrium so that they all share
the same common temperature. The weaker the coupling, the longer the time
required to come to equilibrium. The temperatures in different PF factors in
the product need not be the same as there is no coupling between independent
degrees of freedom. Therefore, such a situation does not have to be considered
here. We only consider the case of an irreducible PF in this work.

The microstates over which the operation $\widehat{\text{Tr}}$ is carried out
in (\ref{dofZ}) are determined by the coupled degrees of freedom, while the
macrostates are determined by the thermodynamic variables $N$, $V$ and the
temperature $T$. Since the degrees of freedom are not part of the macroscopic
description, but the average energy defined below is, it is convenient to
replace the trace operation in (\ref{dofZ}) by a trace only over the energy of
each microstates. We classify different realizations of the degrees of
freedom, i.e. the microstates, according to their energy $E$, so that $W(E)$
is the number of microstates of energy $E$. We can now transform the above PF
in (\ref{dofZ}) into%

\begin{equation}
Z(T)\equiv\text{Tr }W(E)\exp(-\beta E), \label{PF}%
\end{equation}
where Tr now stands for the trace operation over the energy (eigenvalues)
$E\geq E_{0}$ up to its maximum$.$ We also introduce the following extensive
quantities
\begin{equation}
\Omega(T)\equiv\ln Z,\;\;\;F(T)\equiv-T\ln Z; \label{FreeEs}%
\end{equation}
here, $F(T)$ is the conventional Helmholtz free energy.

Since the sign of the entropy is an important issue in the study of glasses,
it is important that the entropy be introduced using the \emph{number} of
microstates $W(E),$ so that $W(E)\geq1$. This requires some kind of
\emph{discretization} of the degrees of freedom \cite{Note2} to count the
microstates. In the following, we will assume that such a discretization has
been carried out so that we always have $W(E)\geq1$ for states that occur in
Nature. This ensures that the corresponding \emph{microstate} entropy
$S(E)\equiv\ln W(E)$\ due to the coupled degrees of freedom\ is non-negative.
Despite this, we will see below that negative entropies can be obtained from
the free energy associated with the SMS at low temperatures, even though the
free energy itself is stable \cite{Note3}. This will force us to conclude that
an ideal glass transition must intervene to avoid the resulting entropy crisis.

The PF in (\ref{PF}) is irreducible. It can happen in some cases that the PF
is a product of several non-trivial PF's corresponding to independent sets of
degrees of freedom, each set containing only coupled degrees of freedom; see
\cite{Note1}. The application of the reality condition to each set requires
that the entropy from each set be non-negative for the system to occur in
Nature. The entropy crisis occurs when any of the entropies from the
independent sets (each containing coupled degrees of freedom) becomes negative
even if the total entropy due to all sets is non-negative.

\subsection{Thermodynamic Limit}

The thermodynamic limit is obtained by taking $N\rightarrow\infty,$ keeping
$v\equiv V/N$ fixed. To make the discussion clear, we will exhibit the
subscript $N$ in various quantities in this subsection. The limit is taken by
considering the sequence formed by
\[
\omega_{N}(T)\equiv(1/N)\Omega_{N}(T),\;\text{where}\;\Omega_{N}(T)\equiv\ln
Z_{N}(T),
\]
for different values of $N$ as $N\rightarrow\infty.$ The volume may be changed
according to $V=vN.$ For proper thermodynamics, the limit of the sequence must
exist, which we denote by $\omega.$

We express the fact that $V/N$ is fixed by stating that $V$ is a homogeneous
function of order 1 in $N.$ Similarly, the existence of the limit $\omega$ is
expressed by stating that $\Omega_{N}(T)\ $or $F_{N}(T)\equiv-T\ln Z_{N} $ are
homogeneous functions of order 1 in $N.$ One can also say that the temperature
$T$ is a homogeneous function of order 0 in $N.$ The average energy $E_{N}(T)$
and the entropy $S_{N}(T)$\ are homogeneous functions of order 1 in $N.$ For
any quantity $Q_{N},$ which is a homogeneous function of order 1 in $N,$ the
ratio $Q_{N}/N$ is a homogeneous function of order 0 in $N$ and must possess a
limit $q$\ as $N\rightarrow\infty.$ We express this fact in the following as
\begin{equation}
Q_{N}/N\thicksim q, \label{HF00}%
\end{equation}
whose significance is as follows:\ for finite $N$, the meaning of (\ref{HF00})
is that the right side may differ from $q$, but the difference vanishes as
$N\rightarrow\infty.$

\textbf{Remark 1}: In the following, whenever we compare different extensive
quantities $Q_{i}$ or different PF's $Z_{i}$, it is implicit that we are
comparing the ratios $Q_{i}/N$ or $\ln Z_{i}/N,$ respectively.

\subsection{Continuous Energy and Entropy Functions for Finite $N$}

The average energy and entropy in the canonical ensemble are continuous
functions of $T$ and are given by%

\begin{subequations}
\begin{align}
\overline{E}(T)  &  \equiv-(\partial\Omega/\partial\beta),\label{AvE}\\
\overline{S}(T)  &  \equiv-(\partial F/\partial T), \label{AvS}%
\end{align}
respectively; we have suppressed the subscript $N$ for simplicity. They should
not be confused with the microstate energy $E$ and entropy $S(E).$ To see this
clearly, we rewrite (\ref{AvS}) as\ $\overline{S}=\ln Z+\overline{E}/T,$ and
exponentiate it. We can use the explicit $T$-dependence of $\overline{E}%
(T)$\ to express $\overline{S}(T)$ as an explicit\ function $S(\overline
{E})\ $of $\overline{E}\equiv\overline{E}(T).$ Introducing $W(\overline
{E})\equiv\exp[S(\overline{E})]$, \ we have
\end{subequations}
\begin{equation}
W(\overline{E})=\text{Tr}\;W(E)e^{-\beta(E-\overline{E})}=W(E_{\text{m}%
})e^{-\beta(E_{\text{m}}-\overline{E})}+\text{Tr}^{^{\prime}}\;W(E)e^{-\beta
(E-\overline{E})}, \label{ContW}%
\end{equation}
where the prime on the trace indicates that some microstate energy
$E=E_{\text{m}}$\ is not allowed in the trace. For a finite system ($N<\infty
$), the allowed microstate energy $E,$ and the microstate entropy
$S(E)\equiv\ln W(E)$\ are discrete, while the average energy\ $\overline{E}$
and the average entropy $\overline{S}$ are continuous functions. Let us
consider the case when the average energy is exactly equal to the microstate
energy $E_{\text{m}}.$ From (\ref{ContW}), we observe that
\begin{equation}
W(\overline{E})\geq W(E_{\text{m}}). \label{ContDis}%
\end{equation}
\ The left side in (\ref{ContDis}) represents the value of the continuous
function $W(\overline{E}=E_{\text{m}}),$\ while the right side is the discrete
quantity $W(E_{\text{m}})$ for the finite system. The difference between them
is due to the last term in (\ref{ContW}), which is non-negative, which
vanishes as $N\rightarrow\infty.$ For finite $N,$\ the microstate entropy
$S(E)$ represent isolated points, which get closer and closer to the
continuous function $S(\overline{E})$ as $N$\ increases, and cover it entirely
as $N\rightarrow\infty.$\ Similarly, $\overline{E}$ represents a continuous
variable, and contains in its range isolated values given by the microstate
energies; the latter cover the entire range of $\overline{E}$ as
$N\rightarrow\infty.$\ 

The continuous function $S(\overline{E})$ is amenable to differentiation,
which is not possible with the discrete set of points $S(E),$ and contains all
the useful physical information. Because of this, there is no harm in
expressing $S(\overline{E})$ simply by $S(E)$; the latter now represents a
continuous function$.$ Similarly, there is no harm in expressing the
continuous variable $\overline{E}$ by $E$. This is a common practice in the
literature. Whenever we need to make a distinction between the discrete values
and the continuous functions, we will speak of the values related to
microstates in the former case, and of average values in the latter case.
Similarly, we use $S(T)$ to express the entropy $\overline{S}(T)\equiv
S(\overline{E})$ as a function of $T$, where $T$ is the temperature at which
the average energy from (\ref{AvE}) is $\overline{E}.$ The use of the
continuous functions enables us to consider the entropy for any value of the
energy, whether it represents the energy of a microstate or not. The
continuous entropy function satisfies the following relation:
\begin{equation}
(\partial S/\partial E)\equiv(\partial S/\partial T)/(\partial E/\partial
T)\equiv1/T. \label{Temp0}%
\end{equation}
\qquad\qquad\ 

Let us first illustrate the above points by a simple example. Consider two
ferromagnetically interacting (interaction strength $-J$) Ising spins in the
absence of any external magnetic field. There are four microstates, two with
$E=-J$ (the lowest possible energy)$,$ and two with $E=J$ (the highest
possible energy)$.$ Thus, there are only two microstate energies, each having
the microstate entropy $\ln2$. On the other hand, the average energy and
entropy are continuous functions of $T$. To see this, we calculate the
canonical PF, $\ $which is $Z\equiv4\cosh K,$ where $K=J/T.$ A simple
calculation for the average energy and entropy yields
\[
E=-J\tanh K,\;\;S=\ln(4\cosh K)-K\tanh K.
\]
We observe that the energy ranges continuously between $-J\ $and $0\ ,$ and
the entropy ranges continuously between $\ln2$ and $\ln4$\ for $T\geq0.$ (The
energy range between $0$ and $J$ corresponds to negative temperatures that we
do not consider here.) The average energy and entropy are $(-J)$ and $\ln2,$
respectively, as $T\rightarrow0,$ and $0$ and $\ln4,$ respectively, as
$T\rightarrow\infty$, as expected. The average energy, and entropy fill
continuously the microstate energy and entropy gap, respectively. The
important point is that \emph{the lowest energy corresponds to the absolute
zero of the temperature.}

We now prove that the entropy slope relation in (\ref{Temp0}) is valid for any
$N$, finite or not. We consider the PF\ in (\ref{PF}) for finite $N.$ We start
with the continuous entropy and energy functions related by $TS=E-F$.
Differentiating on both sides with respect to $T$, we find
\[
S+T(\partial S/\partial T)=(\partial E/\partial T)+\ln Z+T(\partial Z/\partial
T)/Z.
\]
Recognizing that
\[
(\partial Z/\partial T)=(1/T^{2})\text{Tr }EW(E)\exp(-\beta E),
\]
we find that the above equation reduces to
\[
T(\partial S/\partial T)=(\partial E/\partial T),
\]
which proves (\ref{Temp0}) for finite $N$. The limit $N\rightarrow\infty$ is
considered by dividing both sides of $TS=E-F$ by $N$, and carrying out the
same steps.

We now argue that if we replace $W(E)$ by $W(\overline{E})$ in (\ref{PF}) to
define a new PF
\begin{equation}
\overline{Z}(T)\equiv Tr\;W(\overline{E})e^{-\beta\overline{E}},
\label{ContPF0}%
\end{equation}
in which the trace is still over the microstate energies, then using
$\overline{S}=\ln Z+\overline{E}/T,$ and $W(\overline{E})\equiv\exp
(\overline{S})$, we find that
\begin{equation}
\overline{Z}(T)\equiv Z(T)Tr\;1\thicksim Z(T), \label{ContPF}%
\end{equation}
since $Tr\;1$ is the number of distinct microstate energies and satisfies
$\ln(Tr\;1)/N\thicksim0\cite{Guj1}.$ Consequently, $\ln\overline{Z}(T)/N$ and
$\ln Z(T)/N$ are the same in the sense of the above Remark 1. In the
following, we will consider both versions of the PF for macroscopically large
but finite systems, as they are identical in all thermodynamic consequences.
Indeed, it is a common practice in the literature not to make any distinction
at all. Because of this, we will denote both of them by $Z(T) $; we will
indicate the difference whenever needed.

\textbf{Remark 2: }It should be noted that the association of the slope in
(\ref{Temp0}) with the temperature $T$ requires fixing the temperature of the
system. The temperature is fixed from outside, such as by using a heat bath,
and must be \emph{independent} of the size $N$\ of the system, even though
both $\ S$ and $E$ depend on $N$ in accordance with (\ref{HF00}). In this
sense, the significance of $T\equiv$ 1/$(\partial S/\partial E)$\ is that for
any $N$, there exists an $E$ so that (\ref{Temp0}) is always satisfied. $\ $

\subsection{Configurational PF}

In classical statistical mechanics (CSM), the positional degrees of freedom
are independent of the translational (i.e., momentum) degrees of freedom when
the collisions are neglected. In this case, the PF can be written as a product
of two PF's. One of them is $Z_{\text{KE}}(T_{\text{KE}})$ determined by the
momentum degrees of freedom; here $T_{\text{KE}}$ is the temperature
associated with these degrees of freedom and the energy associated with these
degrees of freedom is the kinetic energy of the system. (It should be noted
that the momentum degrees of freedom are also independent of each other so
each momentum degree of freedom can have its own temperature. We will not
worry about this complication.) The other factor is the PF $Z(T),$ known
conventionally as the configurational PF at a temperature $T$. It is
determined only by the positional degrees of freedom for which the energy is
the potential energy in the system. The collisions between particles are
neglected so the two sets of the degrees of freedom have no mechanism to come
to equilibrium. Hence, there is no way for the two temperatures to be the
same. Usually, it is assumed that there exists a weak coupling between the two
degrees of freedom, which is sufficient to bring about eventual equilibrium
between them. Thus, to a good approximation, the above factorization is taken
to be valid in real systems. The important observation is that the value of
$Z_{\text{KE}}$ is independent of the value of the configurational PF $Z(T)$
at this level of the approximation$.$ Consequently, the total entropy
$S_{\text{tot}}(T)$ due to both energies is additive: $S_{\text{tot}}(T)\equiv
S(T)+S_{\text{KE}}(T)$, where $S_{\text{KE}}(T)$ is the entropy due to the
kinetic energy and is\emph{\ independent} of the configurations of the system.
Furthermore, $S_{\text{KE}}(T)$ is the same for all kinds of systems. Thus,
$Z_{\text{KE}}(T)$ is not of any interest when studying any particular system.
Because of this, there is no harm in restricting our attention to the studying
the configurational PF $Z(T).$ In this case, the PF in (\ref{PF}) represents
the configurational PF $Z(T)$ as defined conventionally in CSM so that $S(T)$
will represent the conventional configurational entropy in CSM. The entropy
crisis occurs when $S(T)$ becomes negative, even if $S_{\text{tot}}(T)$ is non-negative.

\subsection{Quantum PF}

In quantum statistical mechanics (QSM), the kinetic energy is an operator and
cannot be separated out from the total PF\cite{Note1}. The role of the
configurational PF is now played by the total PF. The energy $E$ in (\ref{PF})
now represents the eigenvalues of the total (potential+ kinetic) energy, and
$S(T)$ derived from (\ref{PF}) now represents the total entropy, which cannot
be broken into additive terms as was the case in CSM above. Thus, one
\emph{cannot} define the classical configurational entropy in QSM. We can
think of $S(T)$ derived from (\ref{PF}) as the quantum analog of this
classical concept.

The irreducible PF $Z(T)$ in (\ref{PF}) is the general form of the PF valid in
both CSM and QSM, with $S(T)\equiv S(\overline{E})$ equal to the conventional
configurational entropy in CSM, and the total entropy in QSM. From now onward,
we will no longer explicitly distinguish between the classical and quantum
PF's. Our discussion is going to be valid for both cases.

\subsection{Conditions for Equilibrium and Negative Entropy}

For the lowest allowed energy $E_{0},$ we must surely have $W(E_{0})\neq0.$
Assuming $TS(T)\rightarrow0$ as $T\rightarrow0$, we recognize that $E_{0}$
represents not only the Helmholtz free energy but also the energy of the
perfect CR at $T=0$. (We assume that CR has the lowest free energy at low
temperatures.) Since $W(E)$ is non-negative, $Z$ is a sum of positive terms.
Because of this, the probability of every microstate is strictly non-negative.
As a consequence, the following two principles are always satisfied.

(1) \emph{Maximization Principle. }The PF $Z$ must be \emph{maximized} in the
thermodynamic limit $N\rightarrow\infty,$ keeping $V/N$ fixed. The maximum
value of $Z(T)$ corresponds to picking out the maximum term $e^{S-\beta E}$ in
(\ref{PF}). This maximum term corresponds to $E=\overline{E},$ defined above.

(2) \emph{Stability Principle. }The heat capacity, which is given by thy
fluctuations in the energy is non-negative. This remains true even in the
thermodynamic limit.

It should be stressed that the non-negativity of the heat capacity and the
maximization principle only require the positivity of $W(E).$ Thus, both
principles remain valid even if the entropy becomes negative\cite{Note3}.
Stability and reality are two independent aspects of our formalism. This
observation is going to be useful when we discuss the metastable states below.

\section{Stationary Metastable States and Restricted Ensemble}

\subsection{Infinite System}

Conventional statistical mechanics describes equilibrium states, which satisfy
the above two principles of reality and global maximization. For this, it is
necessary that we have $N\rightarrow\infty.$ The existence of a melting
transition, which also requires $N\rightarrow\infty,$ at $T_{\text{M}}$\ means
that the \emph{disordered }equilibrium liquid (EL) phase above $T_{\text{M}}$
and the ordered CR below $T_{\text{M}}$\ correspond to different values of the
order parameter $\rho,$ which is traditionally used to distinguish various
phases of the system, with $\rho=0$ representing the disordered phase and
$\rho\neq0$ the ordered phase CR$.$ (One of our examples below will show
explicitly how the microstates can be divided into the two disjoint classes.)
We assume here for simplicity that there is only one kind of ordered phase.
The extension to the case of many disjoint ordered states with different
non-zero values of $\rho$ is easy to incorporate in the approach. This
distinction in the order parameter is easily made in the case $N\rightarrow
\infty,$ by denoting the free energy per particle ($N\rightarrow\infty$) above
$T_{\text{M}}$ by $f_{\text{dis}}(T),$\ and below $T_{\text{M}}$ by
$f_{\text{ord}}(T),$ from which we can calculate the entropies, and energies
per particle
\begin{equation}
s_{\alpha}(T)\equiv-(\partial f_{\alpha}/\partial T),\;e_{\alpha}%
(T)\equiv-(\partial\beta f_{\alpha}/\partial\beta),\;\;\alpha=\text{dis, ord,}
\label{sealpha}%
\end{equation}
respectively, corresponding to the two states. Due to the limit $N\rightarrow
\infty$, the per particle microstate entropies and continuous entropy
functions are equal at the same $e$. Thus, we can think of the continuous
entropy and energy functions $s_{\alpha}(E),$ and $e_{\alpha}(T) $ for the two states.

The global maximization, which operates when $N\rightarrow\infty,$ is required
to argue that $f_{\text{ord}}(T)$ is the equilibrium state free energy (for
CR) below $T_{\text{M}},$ and $f_{\text{dis}}(T)$ the equilibrium state free
energy (for EL) above $T_{\text{M}}:\;\;\;$%
\begin{subequations}
\begin{align}
f(T) &  \equiv f_{\text{ord}}(T)\;\;\;T<T_{\text{M}},\label{freeenergyalpha}\\
f(T) &  =f_{\text{dis}}(T)\;\;\;T>T_{\text{M.}}\label{freeenergyalpha1}%
\end{align}
Thus, incorporating the global maximization principle will not allow us to
describe SCL. The singularity in the equilibrium free energy per particle
$f(T)$ at $T_{\text{M}}$ in (\ref{freeenergyalpha},\ref{freeenergyalpha1})$,$
which forces it to switch from $f_{\text{dis}}(T)$ to $f_{\text{ord}}(T)\ $at
$T_{\text{M}}$, is a hallmark of the a phase transition. This singularity in
$f(T)$ does not necessarily imply a singularity in either of its two pieces
$f_{\text{dis}}(T)$ and $f_{\text{ord}}(T).$ Both$\ $of them can exist on
either side of $T_{\text{M}}.$ This possible extension is not a consequence of
a mean-field approximation, as our first example will demonstrate. A
consequence of this is that the functions $s_{\alpha}(T),$ and $e_{\alpha}(T)$
also exist on either side of $T_{\text{M}}.$ In the following, we are only
interested in the case in which $s_{\text{dis}}(T),$ and $e_{\text{dis}}%
(T)$\ exist all the way down to $T=0$.

A prescription to describe metastability using the PF\ formalism can now be formulated.

\textbf{Metastability Prescription} We \emph{abandon} the global maximization
principle, and use $f_{\text{dis}}(T)$ to give the free energy of the
metastable disordered phase (supercooled liquid) below $T_{\text{M}}$
and\ $f_{\text{ord}}(T)$ to give the metastable (superheated crystal) state
free energy above $T_{\text{M}}.$ Similarly, $s_{\text{dis}}(T),$
$e_{\text{dis}}(T)$ and $s_{\text{ord}}(T),$ $e_{\text{dis}}(T)$ give the
entropy and energy per particle for the supercooled liquid and superheated
crystal, respectively.

However, in this work, we are only interested in the supercooled liquid.

\subsection{Finite System}

The study of an infinite system allows us to identify ordered ($\rho\neq
0$)\ and disordered ($\rho=0$) microstates and gives their entropy and energy
per particle $s_{\alpha}(T),$ and $e_{\alpha}(T),$ respectively. The
identification is useful to classify each microstate as ordered ($\rho\neq0 $)
or disordered ($\rho=0$). This does not imply that it is feasible to count
these microstates for an infinite system in which there are infinite
microstates. For a complete analysis, we need to be able to count in principle
the microstates and classify them. For this, we need to consider the case of
finite but large $N.$ In this case, the continuous entropy and energy
functions for the entire system are $S_{\alpha}(T)\thicksim Ns_{\alpha}(T),$
and $E_{\alpha}(T)\thicksim Ne_{\alpha}(T).$ We also consider the continuous
functions $S_{\alpha}(E),$\ and $W_{\alpha}(E)\equiv\exp[S_{\alpha}(E)]$
which, as we show later, \emph{exist} for \emph{all} energies $E\geq E_{0}.$
As said earlier, the continuous function $W_{\alpha}(E)$ for
microstate\ energy $E$\ is close to the number of ordered and disordered
microstates of microstate\ energy $E$; see (\ref{ContDis}).

The total number of microstates $W(E)$ with microstate energy $E$\ can be
written as a sum of the number of microstates $W_{\text{dis}}(E)$ consistent
with $\rho=0$, and the number of microstates $W_{\text{ord}}(E)$ consistent
with $\rho\neq0,$ so that
\end{subequations}
\begin{equation}
W(E)\equiv W_{\text{ord}}(E)+W_{\text{dis}}(E). \label{Partition}%
\end{equation}
(Presence of other configurations will not affect our argument.)\ While
$W_{\text{ord}}(E)$ certainly exists for microstate energies starting from
$E=E_{0},$ there is no guarantee that $W_{\text{dis}}(E)$ also exists near
$E=E_{0}$. Most probably, microstate $W_{\text{dis}}(E)\geq1$ does not
continue all the way down to $E=E_{0}$. If it did, the energy of the
disordered phase at absolute zero would be $E_{0},$ the same as that of CR.
This would most certainly imply that they would coexist at $T=0$, each having
the same volume; recall that we are considering a fixed volume ensemble$.$%
\ While there is no thermodynamic argument against it, it does not seem to be
the case normally. Usually, the most stable state at $T=0$ is that of a
crystal. Moreover, it is an experimental fact\cite{Kauzmann} that all glasses
have much higher energies or enthalpies compared to their crystalline forms at
low temperatures; see (Expt2) above. Anticipating that the ideal glass, which
is the stationary limit of all observed glasses, should have its energy higher
than $E_{0},$ we conclude that the lowest possible energy $E_{\text{K}}$\ for
the disordered state is strictly larger than $E_{0}.$ In other words, the
microstate number $W_{\text{dis}}(E)$ has the following property:
\begin{subequations}
\begin{align}
W_{\text{dis}}(E)  &  =0\;\;\;\text{for \ \ }E<E_{\text{K}},\label{Wdis0}\\
W_{\text{dis}}(E)  &  \geq1\ \text{ \ \ for \ \ }E\geq E_{\text{K}}.
\label{Wdis}%
\end{align}
On the other hand, we are only interested in considering the case when
$s_{\text{dis}}(T)$ exists for all $T\geq0.$ Thus, $S_{\text{dis}}(T)$ exists
for all $T\geq0.$ We will show later that the temperature $T=T_{\text{K}%
}=1/(\partial S_{\text{dis}}/\partial E)$ at $E_{\text{K}}>E_{0}$
corresponding to the disordered phase is strictly positive. Thus,
$S_{\text{dis}}(T)$ [or $S_{\text{dis}}(E)$] must be negative below
$T=T_{\text{K}}$ [or $E=E_{\text{K}}].$ We will, in fact, see that
$S_{\text{dis}}(E)$ continues all the way down to $E_{0}.$

\subsection{Restricted Ensemble}

We continue with finite but large $N$. Using the continuous forms of
$W_{\text{ord}}(E)$ and $W_{\text{dis}}(E)$, we introduce the following
\emph{restricted ensemble }approach to describe metastable
states\cite{Penrose}. We follow (\ref{ContPF0}) and we introduce two new PF's
using (\ref{PF}) by replacing continuous $W(E)\ $by continuous $W_{\text{ord}%
}(E)$ and $W_{\text{dis}}(E),$ respectively:
\end{subequations}
\begin{equation}
Z_{\alpha}(T)\equiv TrW_{\alpha}(E)\exp(-\beta E),\;\;\;\alpha=\text{dis,
ord,} \label{PFalpha}%
\end{equation}
and the corresponding free energies
\begin{equation}
F_{\alpha}(T)\equiv-T\ln Z_{\alpha}(T),\;\;\;\alpha=\text{dis, ord.}
\label{Falpha}%
\end{equation}
As $N\rightarrow\infty$, $F_{\alpha}(T)/N\rightarrow f_{\alpha}(T).$

Above the melting temperature $T_{\text{M}},$ $Z_{\text{dis}}$
$>$%
$Z_{\text{ord}},$ and below $T_{\text{M}},$ $Z_{\text{dis}}$%
$<$%
$Z_{\text{ord}}$. The use of the maximization principle describes CR and EL
collectively for thermodynamically large $N$:
\begin{subequations}
\begin{align}
Z_{\text{dis}}(T)  &  >Z_{\text{ord}}(T),\;\;\;\;\;T>T_{\text{M}}%
,\label{PFEL}\\
Z_{\text{ord}}(T)  &  >Z_{\text{dis}}(T),\;\;\;\;\;T<T_{\text{M}},\;
\label{PFCR}%
\end{align}
so that
\begin{align}
F_{\text{ord}}(T)  &  >F_{\text{dis}}(T),\;\;\;\;\;T>T_{\text{M}%
},\label{FreeEEL}\\
F_{\text{dis}}(T)  &  >F_{\text{ord}}(T),\;\;\;\;\;T<T_{\text{M}}.
\label{FreeECR}%
\end{align}

For finite $N$, there cannot be any singularity at $T_{\text{M}},$ so that
$F_{\text{dis}}(T)$ or $F_{\text{ord}}(T)$ has no singularity of its own at
$T_{\text{M}}$, i.e., they exist on both sides of $T_{\text{M}}.$\ We now
describe SMS by the PF\ $Z_{\text{dis}}(T)$ below $T_{\text{M}}.$ The
continuation necessarily yields $\rho=0$ also for SCL.

The entropy functions $S_{\alpha}(E)$ are shown in Fig. 1: OHAB represents
$S_{\text{ord}}(E)$, and KCO$^{^{\prime}}$AD represents $S_{\text{dis}}(E)$.
The entropy as a function of $E$ must be thought of as the entropy in the
microcanonical ensemble\cite{Guj1}, which must be maximum for the equilibrium
state. Since a SMS is not an equilibrium state in the unrestricted ensemble,
its entropy at some $E$ \textit{cannot} exceed the entropy of the
corresponding equilibrium state at the same $E$. A consequence of this entropy
condition is that the free energy $F_{\text{dis}}(T)$ of SMS\ \emph{cannot} be
lower than the free energy $F_{\text{ord}}(T)$ of CR at the same temperature
$T.$ This explains the form of the entropy and free energy in Fig. 1. The two
free energies are sown in the inset by OAB representing $F_{\text{ord}}(T)$,
and OKCO$^{^{\prime}}$D representing $F_{\text{dis}}(T)$. The form of the
entropy and free energy is also supported by all known
observations$\cite{Kauzmann,Goldstein}$, exact \cite{GujCorsi,GujRC,Derrida}
and numerical\cite{Sciortino} calculations, and from the arguments and the
calculations presented below. We note that
\end{subequations}
\begin{subequations}
\begin{align}
S_{\text{ord}}(E)  &  <S_{\text{dis}}(E),\;\;\;\;\;E>E_{\text{M}%
},\label{CompS}\\
S_{\text{dis}}(E)  &  <S_{\text{ord}}(E),\;\;\;\;\;E<E_{\text{M}},
\label{CompS1}%
\end{align}
where $E_{\text{M}}$\ is the energy at A where $S_{\text{ord}}%
(E)=S_{\text{dis}}(E);$ see Fig. 1 $.$ The SMS corresponding to the stationary
SCL is defined by the branch KCAH$^{^{\prime}}$. Similarly, superheated CR is
defined by the branch HAB. We note that the entropy $S_{\text{dis}}$ of the
metastable branch goes to zero at $T_{\text{K}}$%
$>$%
0. This behavior is supported by our rigorous proof and by two exact
calculations in the paper. However, we will allow the possibility in our
discussion below that the lowest energy for $S_{\text{dis}}$\ is $E_{\text{D}%
},$ and that $S_{\text{dis}}(E_{\text{D}})>0.$ We also show the entropy
function for a non-stationary metastable state by GF in Fig. 1, assuming that
crystallization is forbidden. We do not consider the non-stationary state GF anymore.

\subsection{A Useful Theorem}

We now prove an extremely useful theorem.\qquad

\begin{theorem}
Since $E_{\text{K}}>E_{0},$ the free energy $F_{dis}(T_{\text{eq}})$ at
O$^{^{\prime}}$equals the free energy $F_{\text{ord}}(T=0)=E_{0}$ at O, where
$T_{\text{eq}}$ is the inverse of the slope of the line OO$^{^{\prime}}$
touching the entropy function $S_{\text{dis}}(E),$ which vanishes at
$E=E_{\text{K}};$ see point O$^{^{\prime}\text{ }}$in the inset in Fig. 1.
\end{theorem}

The proof is very simple. The slope 1/$T_{\text{eq}}$ of OO$^{^{\prime}}$is
given by
\end{subequations}
\[
1/T_{\text{eq}}=S_{\text{dis}}(E_{\text{O}^{^{\prime}}})/(E_{\text{O}%
^{^{\prime}}}-E_{0}),
\]
where $E_{\text{O}^{^{\prime}}}$ is the energy at O$^{^{\prime}}$. Thus,
\[
E_{0}=E_{\text{O}^{^{\prime}}}-T_{\text{eq}}S_{\text{dis}}(E_{\text{O}%
^{^{\prime}}}).
\]
Since the slope of $S_{\text{dis}}(E_{\text{O}^{^{\prime}}})$ at
$E_{\text{O}^{^{\prime}}}$ is $1/T_{\text{eq}},$ the right side represents the
free energy $F_{\text{dis}}(T_{\text{eq}})$ of the SMS at O$^{^{\prime}}.$ The
left side represents the free energy of CR at $T=0.$ This proves the theorem.

It should be stressed that the proof does not use the vanishing of
$S_{\text{dis}}(E_{\text{K}})$. Thus, the equality $F_{\text{dis}%
}(T_{\text{eq}})=E_{0} $ is also valid if $S_{\text{dis}}(E_{\text{D}})>0.$
The proof also does not depend on the entropy slope at $E_{\text{D}}\ $or
$E_{\text{K}}.$ The proof only requires that this slope be larger than or
equal to 1/$T_{\text{eq}}.$

We now consider the behavior of $Z_{\text{dis}}(T).$\ Since the continuous
entropy $S_{\text{dis}}(E)$ is real, we have $W_{\text{dis}}(E)>0.$ Thus, we
observe that $Z_{\text{dis}}(T)$ is a sum of positive terms. Because of this,
$Z_{\text{dis}}(T)$ also satisfies the maximization and stability principles,
just as $Z(T)$ does. The only difference is that they are only valid in the
\emph{restricted ensemble} of the disordered microstates corresponding to
$\rho=0.$

The \textit{reality} condition \textit{cannot} be violated, even for a SMS if
it is observable. Therefore, its violation ($S_{\text{dis}}(T)<$ $0$ for
$T<T_{\text{K}}>0$) implies that SMS \emph{cannot} exist in Nature when the
violation occurs, in which case the SMS associated with the SCL must be
replaced by a new state, commonly known as the \textit{ideal glass state}
below $T_{\text{K}},$ whose energy at $T\leq T_{\text{K}}>0$ is $E_{\text{K}%
}>E_{0}\cite{Kauzmann,Goldstein}$. The transition between the two states is
called the ideal glass transition.

\section{Finite Entropy Slope of the Disordered Phase}

\begin{theorem}
The slope of the entropy function $S_{\text{dis}}$ at $E/E_{0}>1$ must be finite.
\end{theorem}

We again consider finite and infinite $N$ separately.

\subsection{Finite $N$}

We consider the average entropy functions $S_{\alpha}(E)$ for fixed $V$,$N$;
see Fig. 1, where they are shown schematically as functions of $E.$ According
to (\ref{Temp0}) $\qquad\qquad$%
\begin{equation}
(\partial S/\partial E)\equiv1/T,\;\;\;(\partial S_{\alpha}/\partial
E)\equiv1/T_{\alpha}, \label{Temp}%
\end{equation}
where $1/T$, $1/T_{\alpha}$ represent the inverse temperatures corresponding
to $S$, $S_{\alpha}$, respectively$.$ By introducing the entropy and energy
densities
\begin{equation}
s_{N}\equiv S/N,\;s_{\alpha,N}\equiv S_{\alpha}/N,\,\;e_{N}\equiv E/N,
\label{TempN}%
\end{equation}
we can rewrite (\ref{Temp}) that will be useful in taking the thermodynamic
limit below. For finite $N,$ $s_{N},s_{\alpha,N},$\ and $e_{N}$ are
homogeneous functions of order 0; see (\ref{HF00}).

The behavior of the slope in (\ref{Temp}) for the disordered phase is critical
in understanding what happens at K, the Kauzmann point (where $S_{\text{dis}%
}=0)$. There are two distinct possibilities. The slope at K is either finite,
as shown explicitly in Fig. 1 and consistent with our rigorous analysis, or
infinite. The former corresponds to a positive $T_{\text{K}},$ while the
latter corresponds to an absence of a Kauzmann paradox ($T_{\text{K}}=0$).
Almost all explicit calculations\ show the former as the usual behavior, two
of which are presented below.

To proceed further, we consider (\ref{Partition}) for any microstate energy.
From what has been said earlier about the continuous function $S(E),$\ we
extend (\ref{Partition}) to any allowed continuous energy so that it can be
differentiated. Taking the derivative with respect to $E$ on both sides, we
find that%

\[
W(\partial S/\partial E)\equiv W_{\text{ord}}(\partial S_{\text{ord}}/\partial
E)+X_{\text{dis}}(\partial S_{\text{dis}}/\partial E).
\]
We introduce the ratios $X_{\alpha}\equiv W_{\alpha}/W$ ($0\leq X_{\alpha}%
\leq1$) and rewrite the above equation as
\begin{equation}
1/T\equiv X_{\text{ord}}/T_{\text{ord}}+X_{\text{dis}}/T_{\text{dis}}\text{.}
\label{slope}%
\end{equation}
For $T>T_{\text{M}},$\ $X_{\text{ord}}\approx\exp(S_{\text{ord}}%
-S_{\text{dis}})<<1,$\ while $X_{\text{dis}}\approx1.$ For $T<T_{\text{M}}%
,$\ $X_{\text{dis}}\approx\exp(S_{\text{dis}}-S_{\text{ord}})<<1,$\ while
$X_{\text{ord}}\approx1.$ Let us apply this relation to the crystal at\ a
temperature $T_{\text{M}}>T>0$\ at which the average CR energy is
$E>E_{\text{0}}$. It is clear that for any \textit{finite} $N$, no matter how
large, $X_{\text{dis}}>0.$\ At the same time, the temperature $T$ must be
close to $T_{\text{ord}}$ of the CR, which is strictly positive, since
$E>E_{0}.$ This is possible only if
\begin{equation}
T_{\text{dis}}>0\;\;\text{ for\ \ \ }E/E_{0}>1. \label{FiniteSlope}%
\end{equation}
Thus, as long as $E>E_{0}$, the temperature of the disordered phase at
$E$\ must be $positive$. This proves the theorem for finite $N$.

\subsection{Infinite $N$}

Let us now take the limit $N\rightarrow\infty.$\ We only consider the
interesting range $T_{\text{M}}>T>0.$ The values of $E$\ and $S$ keep changing
with $N$ \ for a given $T$. \ Thus, it is convenient to consider the sequences
\{$e_{N}\},$ and \{$s_{\alpha,N}\}$ for different $N$, so \ that $(\partial
s_{\alpha,N}/\partial e_{N})=1/T_{\text{dis}}>0.$ We now consider
(\ref{slope}) for larger and larger $N$, such that $T_{\text{dis}}>0\;$ is
kept fixed. From what was said earlier, it is clear that $X_{\text{dis}%
}\rightarrow0$ from above$,$ so that $T\rightarrow T_{\text{ord}}$ from
below.\ But at every step of the limit $N\rightarrow\infty$,
(\ref{FiniteSlope}) remains valid. Thus, $T_{\text{dis}}>0$ for \ $E/E_{0}>1$
even when $N\rightarrow\infty.$

The above proof of (\ref{FiniteSlope}) neither requires nor shows that the
entropy $S_{\text{dis}}(E)=0$ for $E>E_{0}$. It also does not depend on the
sign of $S_{\text{dis}}(E)$\ as long as it remains differentiable. It only
requires $W_{\alpha}$\ to be non-negative for $T_{\alpha}$\ to be \emph{real}.
Thus, it does not directly prove the existence of a Kauzmann point below which
the entropy crisis would occur ($W_{\text{dis}}<1$). For this, we need to show
that the entropy of the disordered phase becomes zero at some $E_{\text{K}%
}>E_{0}.$ Then the above theorem proves that this occurs at a positive
temperature. Furthermore, it shows that the disordered phase can reach the
absolute zero only if $E=E_{0}.$ The proof that an entropy crisis does occur
in general at a positive temperature\ is given in the following section.

\section{Free Energy at $T=0$}

We now prove the following theorem.

\begin{theorem}
The free energy $F_{\alpha}$ of all stable phases, mathematically continued or
not, are equal in that $F_{\alpha}/E_{0}\rightarrow1$, provided $Ts_{\alpha}$
$\rightarrow0$ as $T\rightarrow0.$ Their entropies $s_{\alpha},$ however, may
be different$.$
\end{theorem}

The proof will require considering finite $N$ and then taking the
thermodynamic limit later just as above.

\subsection{Finite $N$}

Consider finite $N$. From (\ref{Wdis0}), we conclude that microscopic entropy
$S_{\text{dis}}(E)$ does not exist (as a bounded quantity) for $E<E_{\text{K}%
}.$ On the other hand, it must be close to the continuous function
$S_{\text{dis}}(E)$ exists only for $E\geq E_{\text{K}},$ and vanishes at
$E_{\text{K}}$, as shown in Fig. 1. This is consistent with \ (\ref{Wdis}).
From Theorem 2, we conclude that the temperature $T_{\text{K}}$\ corresponding
to $S_{\text{dis}}(E)$ at $E_{\text{K}}$\ is positive. Moreover, since $N$ is
finite, $S_{\text{dis}}(E)$ is not singular at $E_{\text{K}}.$ (Recall again
that the singularity can only appear in the limit $N\rightarrow\infty.$ Our
examples will show that the entropy per site $s_{\text{dis}}(E)$\ remains
non-singular even in the thermodynamic limit.) Then, it can be continued
\emph{mathematically} as a real function below $E_{\text{K}}$ all the way down
to $E_{0}.$ But the continuation of $S_{\text{dis}}(E)$ for $E<E_{\text{K}},$
which must necessarily lead to negative $S_{\text{dis}}(E)$ there$,$ is
certainly not close to the microscopic entropy\ $S_{\text{dis}}(E)$
$(=-\infty)$. \ Thus, the mathematical continuation of$\ S_{\text{dis}}(E)$
below $E_{\text{K}}$ will most certainly not represent the physics correctly.
This is not surprising in view of the fact that it violates the reality
condition. Despite this, the continuous entropy is still useful in the
investigation, and we will allow it to become negative below $E_{\text{K}},$
as it allows us to draw a very important conclusion, as we demonstrate below.
Corresponding, we will allow $W_{\text{dis}}(E)<1$ under continuation.

Due to the mathematical continuation of $W_{\text{dis}}(E),$ both PF's
$Z_{\alpha}(T),$ $\alpha=$dis, ord, contain all energies from $E_{0}$ upwards,
so that both can be investigated in a unified fashion. As long as $W_{\alpha
}(E)\geq0$, $Z_{\alpha}(T)$ is a sum of positive terms, so that $\Omega
_{\alpha}(T)\equiv\ln Z_{\alpha}(T)$ satisfies stability principle. For
example, the free energy $F_{\text{dis}}(T)$ is a concave function (a function
which always lies above the line joining any two points on it) of $T$, as
shown in the inset in Fig. 1 by OKCD. We take out the term corresponding to
$E=E_{0}$ from the trace operation, and express
\begin{equation}
Z_{\alpha}(T)=W_{\alpha}(E_{0})e^{-\beta E_{0}}[1+Z_{\alpha}^{^{\prime}}(T)],
\label{PF1}%
\end{equation}
where we have introduced a new quantity
\begin{equation}
Z_{\alpha}^{^{\prime}}(T)\equiv Tr^{^{\prime}}[W_{\alpha}(E)/W_{\alpha}%
(E_{0})]e^{-\beta(E-E_{0})}, \label{PF2}%
\end{equation}
in which the trace operation $Tr^{^{\prime}}$ is restricted to all $E>E_{0}. $
It is assumed that $W_{\alpha}(E_{0})>0.$ Since $E-E_{0}>0,$\ we note that
$e^{-\beta(E-E_{0})}\rightarrow0$ as $T\rightarrow0.$ For finite $N$,
$W_{\alpha}(E)$\ is a bounded quantity, and so is the ratio $W_{\alpha
}(E)/W_{\alpha}(E_{0})$. Hence, the product under the trace in (\ref{PF2})
vanishes, and so does $Z_{\alpha}^{^{\prime}}(T)$ as $T\rightarrow0.$ We
finally have
\[
Z_{\alpha}(T)\rightarrow W_{\alpha}(E_{0})e^{-\beta E_{0}}\;\text{as
}T\rightarrow0.
\]
We thus find that $F_{\alpha}(T)\rightarrow E_{0}-TS_{\alpha}(E_{0})$ as
$T\rightarrow0.$ From the boundedness of $W_{\alpha}(E_{0})>0$, we also
conclude that $TS_{\alpha}$ $\rightarrow0$ as $T\rightarrow0;$ hence,
\begin{equation}
F_{\alpha}/E_{0}\rightarrow1\;\;\;\text{as }T\rightarrow0 \label{Fzero}%
\end{equation}
for CR and SMS.

\subsection{Infinite $N$}

We recognize that both $F_{\alpha}$ and $E_{0}$ are homogeneous functions of
order 1 in $N.$\ Thus, their ratio $F_{\alpha}/E_{0}$ is a homogeneous
functions of order 0 in $N.$ Consequently, we can take the thermodynamic limit
$N\rightarrow\infty$ by exploiting (\ref{HF00}) without altering the
conclusion $F_{\alpha}/E_{0}\rightarrow1$ as $T\rightarrow0.$ Thus, we finally
conclude that \textit{both }CR\textit{\ and }SMS\textit{\ have identical
energies }$E_{0}$\textit{\ at absolute zero }in the sense that the energy per
particle is the same in both states$.$ However, while the entropy of
CR\ vanishes, that of the mathematically continued SMS has a negative entropy
at $T=0$. Thus, they are \emph{not} identical.

\begin{corollary}
The free energy $F_{\text{dis}}(T),$ mathematically continued to $T=0,$ must
have a maximum at the Kauzmann point $T=T_{\text{K}}$ at which the entropy vanishes.
\end{corollary}

From Theorem 1, we know that $F_{\text{dis}}(T)$=$E_{0}$ at $T=$
$T_{\text{eq}}>0;$ see point O$^{^{\prime}\text{ }}$in Fig. 1. From Theorem 3,
we also know that $F_{\text{dis}}(T)$=$E_{0}$ at $T=0.$ Because of the
non-negative heat capacity $[\partial^{2}F_{\text{dis}}/\partial T^{2}<0],$
$F_{\text{dis}}(T)$\ is a concave function of $T$. Thus, the mathematically
continued$\ F_{\text{dis}}(T)$ must have a maximum between the range
$(0,T_{\text{eq}})$ as shown in the inset in Fig. 1. This maximum at
$T=T_{\text{K}}$\ corresponds to the Kauzmann point at which the entropy vanishes.

Note from Fig. 1, see OO$^{^{\prime}},$ that the equality $F_{\text{dis}%
}(T)=E_{0}$\ at $T=T_{\text{eq}}$\ is independent of the way $S_{\text{dis}}%
$\ approaches zero at $E_{\text{K}}.$ In particular, it does not depend on
whether there is a singularity in $S_{\text{dis}}$ at $E_{\text{K}},$ or
whether $S_{\text{dis}}$\ approaches $E_{\text{K}}$ with a finite slope$.$
Thus, the existence of a maximum in $F_{\text{dis}}(T)$ at $T=T_{\text{K}}$ is
not a consequence of Theorem 1 alone$.$ We need the additional result of
Theorem 3.$\ $

The SMS we have defined mathematically between $T=0$ and $T=T_{\text{K}}$
$\emph{cannot}$ represent a state of any real system because of the negative
entropy, and must be replaced by the ideal glass, which is a state that has
zero entropy and \ a constant energy $E_{\text{K}}$ and represents the
stationary limit of the glassy states observed in experiments. Consequently,
this stationary limit of the glassy state is different not only from the CR,
but also from the SMS near absolute zero. As SMS is cooled to $T=T_{\text{K}%
},$ where it has the energy $E_{\text{K}}>E_{0},$ it turns into the ideal glass.

This completes the proof of the generality of our proposed mechanism. In CSM,
$S$ represents the classical configurational entropy, and in QSM, it
represents the total entropy\cite{Note1}. Thus, $T_{\text{K}}$ corresponds to
vanishing of different entropies in the two cases. Any attempt to estimate the
classical configurational entropy in QSM, where it has no meaning, will
require some sort of approximation, which we do not consider here.
\qquad\qquad\qquad\qquad\qquad\qquad\qquad\qquad\qquad\qquad\qquad\qquad
\qquad\qquad\ \ \ \ \ 

The discussion also establishes that the Kauzmann point in the disordered
phase exists only when there exists another equilibrium state; otherwise,
there will be no partitioning and, therefore, there will be no Kauzmann point.
This is most clearly seen in the first example given below.

\section{Exact Model Calculations}

We consider two CSM models in which we obtain positive Kauzmann temperature.
The calculations are carried out exactly. The first example also shows how
$\rho$ is used to distinguish different phases, while the second example shows
that frustration is not necessary for the glass transition.

\subsection{One-dimensional Axis Spin Model}

We now consider a one-dimensional axis spin model, which contains
$m$-component spins $\mathbf{S}_{i}$ located at site $i$ of the
one-dimensional lattice of $N$ sites, with periodic boundary condition
($\mathbf{S}_{N+1}$ =$\mathbf{S}_{1}$). Each spin can point along or against
the axes (labeled $1\leq k\leq m$) of an $m$-dimensional spin space and is of
length $\sqrt{m}:$ $\mathbf{S=(}0,0,..,\pm\sqrt{m},0,..0)$. The spins interact
via a ferromagnetic nearest-neighbor interaction energy ($-J$), with $K\equiv
J/T>0.$ The energy of the interaction is given by
\[
E=-J\sum_{i=1...N}\mathbf{S}_{i}\cdot\mathbf{S}_{i+1}.
\]
The PF is given by
\begin{equation}
Z_{N}(K,m)\equiv\left(  \frac{1}{2m}\right)  ^{N}\sum_{\;}\exp(-\beta
E)=\left(  \frac{1}{2m}\right)  ^{N}Tr\;\widehat{\text{T}}^{N}, \label{AxisPF}%
\end{equation}
where the first sum is over the $(2m)^{N}$ spin states of the $N$ spins and
$\widehat{\text{T}}\equiv\exp(K\mathbf{S\cdot S}^{\prime})$ is the transfer
matrix between two neighboring spins. The transfer matrix has the eigenvalues
$[x\equiv\exp(Km)]$
\begin{equation}
\lambda_{\text{dis}}=x+1/x+2(m-1),\;\lambda_{\text{ord}}=x-1/x,\;\lambda
=x+1/x-2, \label{Eigen}%
\end{equation}
that are 1-fold, $m$-fold, and ($m-1$)-fold, respectively\cite{GujAxis}.

We follow de Gennes\cite{deGennes,Gujn0} and provide an alternative and very
useful interpretation of the above spin model in terms of a polymer system, in
which each polymer has multiple bonds and loops. The valence at each site in a
polymer must be even. (The presence of a magnetic field will allow odd
valencies, which we do not consider here.) The high-temperature expansion of
the PF, which is given by%

\begin{equation}
Z_{N}(K,m)=\sum K^{B}m^{L}, \label{PolyPF}%
\end{equation}
describes such a polymer system, with $K\geq0$, and $m$ denoting the activity
of a bond and the activity for a loop, respectively, and $B\;$and$\ L$
denoting the number of bonds and the number of loops, respectively\cite{Gujn0}%
. The empty sites represent solvent particles. The number of polymers and the
number of bonds and loops in each polymer are not fixed and vary according to
thermodynamics. In addition, there is no interaction between polymers, and
between polymers and solvent particles, so that the polymer system in
(\ref{PolyPF}) is an \emph{athermal} solution. The temperature $T$ of the spin
system does not represent the temperature in the polymer problem, as is well
known\cite{deGennes,Gujn0}. As we will see below, small $x$ corresponds to
high temperatures where the disordered phase is present, and large $x$
corresponds to low temperatures where the ordered and possible SMS phases are
present. Thus, decreasing $T$ amounts to going towards the region where the
ordered and metastable disordered phases are present. Let $\omega$ denote the
limiting value as $N\rightarrow\infty$\ of
\begin{equation}
\omega\equiv(1/N)\ln Z_{N}(K,m)+\ln(2m), \label{AxisFE}%
\end{equation}
where we have added an uninteresting constant to get rid of the prefactor in
(\ref{AxisPF}). This is done because the number of microstates appears within
the summation in the spin model PF in (\ref{AxisPF}). Thus, the inclusion of
the prefactor will make the microstate entropy negative. The prefactor is,
however, required for the polymer mapping.

The importance of the polymer mapping is that we can take\ $m\geq0$ to be a
real number, even though non-integer $m$ makes no sense for a physical spin.
Thus, for non-integer values of $m$, only the polymer system represents a
physical system. For $m=1,$ the axis model reduces to the Ising model, while
for $m\rightarrow0$, it reduces to the a model of linear chains with no loops
\cite{deGennes,Gujn0}. The eigenvalue $\lambda_{\text{dis}}$ is dominant at
high temperatures for all $m\geq0$\ and describes the disordered phase. Its
eigenvector is
\[
\left\langle \xi_{\text{dis}}\right|  =\sum_{i}\left\langle i\right|
/\sqrt{2m},
\]
where $\left\langle 2k\right|  $ ( or $\left\langle 2k+1\right|  $) denotes
the single-spin state in which the spin points along the positive (or
negative) $k$-th spin-axis. It has the correct symmetry to give zero
magnetization ($\rho=0$). For $m\geq1,$ $\lambda_{\text{dis}}$ remains the
dominant\ eigenvalue at all temperatures $T\geq0$. For $0\leq$ $m<1,$ the
situation changes and $\lambda_{\text{ord}}$ becomes dominant\ at low
temperatures $T<T_{\text{c}},$ or [$x\geq x_{\text{c}}=1/(1-m)$] where
$T_{\text{c}}$\ is determined by the critical value $x_{\text{c}}\equiv
\exp(Jm/T_{\text{c}});$ there is a phase transition at $T_{\text{c}}$. The
corresponding eigenvectors are given by the combinations
\[
\left\langle \xi_{\text{ord}}^{(k+1)}\right|  =[\left\langle 2k\right|
-\left\langle 2k+1\right|  ]/\sqrt{2},\;k=0,2,..,m-1,
\]
which are orthogonal to $\left\langle \xi_{\text{dis}}\right|  $, as can be
easily checked$.$ These eigenvectors have the symmetry to ensure $\rho\neq0 $.
The remaining eigenvalue $\lambda$ is $(m-1)$-fold degenerate with
eigenvectors
\[
\left\langle \xi^{(k+1)}\right|  =[\left\langle 2k\right|  +\left\langle
2k+1\right|  -(\left\langle 2k+2\right|  -\left\langle 2k+3\right|
)]/\sqrt{4},\;k=0,2,..,m-2.
\]
For $m>0,$ this eigenvalue is never dominant. For $m\rightarrow0,$ it becomes
degenerate with $\lambda_{\text{dis}}.$ Since the degeneracy plays no role in
the thermodynamic limit, there is no need to consider this eigenvalue
separately for $m\geq0.$

We now consider the limit $N\rightarrow\infty.$ The adimensional free energy
per site, which represents the osmotic pressure\cite{GujRC,GujEqState}, of the
high-temperature equilibrium phase is $\omega_{\text{dis}}(T)\equiv\ln
(\lambda_{\text{dis}}).$ It can be continued all the way down to $T=0,$ even
though the equilibrium osmotic pressure has a singularity at $x_{\text{c}}.$
Similarly, $\omega_{\text{ord}}(T)\equiv\ln(\lambda_{\text{ord}})$ related to
the low-temperature equilibrium phase can be continued all the way up to
$T\rightarrow\infty.$ To calculate the entropy density, we proceed as follows.
The bond and loop densities are given by
\begin{equation}
\phi_{\text{B}}\equiv\partial\omega/\partial\ln K,\text{\ \ \ \ }%
\phi_{\text{L}}\equiv\partial\omega/\partial\ln m, \label{density}%
\end{equation}
which are needed to calculate the entropy per site of the polymer system
\[
s^{(\text{P)}}=\omega-\phi_{\text{B}}\ln K-\phi_{\text{L}}\ln m;
\]
the superscript is to indicate that it is the polymer system entropy, and is
different from the spin system entropy $s^{(\text{S)}}=\partial T\omega
/\partial T.$ If we define $\omega$ without the last term in (\ref{AxisFE}),
then $\phi_{\text{L}}$ and $s^{(\text{P)}}$\ must be replaced by
$(\phi_{\text{L}}-1)$ and $(s^{(\text{P)}}-\ln2),$ respectively. This will not
affect any of the conclusions below.

In the following, we will be only interested in the polymer entropy. The
proper stability requirements for the polymer system are
\begin{equation}
(\partial\phi_{\text{B}}/\partial\ln K)\geq0,(\partial\phi_{\text{L}}%
/\partial\ln m)\geq0, \label{PolyStbl}%
\end{equation}
as can easily be seen from (\ref{AxisPF}), and must be satisfied even for SMS.
They replace the positivity of the heat capacity of the spin system, which no
longer represents a physical spin system for $0\leq$ $m<1.$ It is easy to see
from the definition of $s_{\text{dis}}^{(\text{P)}}$ that $(\partial
s_{\text{dis}}^{(\text{P)}}/\partial T)_{m}$ need not be positive, even if the
conditions in (\ref{PolyStbl}) are satisfied.

It should be noted that the Theorem 3 was for a canonical PF, whereas the
polymer PF in (\ref{PolyPF}) is an athermal grand canonical PF. The proof of
the theorem can be easily extended to this or any other ensemble with similar
results. Here, we will instead give a direct demonstration of the theorem. For
this, we compute $\omega$\ as $K\rightarrow\infty$ ($T\rightarrow0)$ for the
two eigenvalues $\lambda_{\text{dis}}$ and $\lambda_{\text{ord}}$. From
(\ref{density}), it is easy to see that $\phi_{\text{B}}\rightarrow mK$ for
both states as $T\rightarrow0.$ Thus, using $\omega=s^{(\text{P)}}%
+\phi_{\text{B}}\ln K+\phi_{\text{L}}\ln m,$ we have
\begin{equation}
\omega_{\text{dis}}(T)/\omega_{\text{ord}}(T)\rightarrow1\;\text{\ as
\ }T\rightarrow0. \label{EqFE}%
\end{equation}
This is in accordance with (\ref{Fzero}). This means that if the eigenvalue
$\lambda_{\text{dis}}$ is taken to represent the metastable phase above
$x_{\text{c}}$, its osmotic pressure must become equal to that of the
equilibrium phase (described by the eigenvalue $\lambda_{\text{ord}}$) at
absolute zero, in conformity with Theorem 3. We take $\omega_{\text{dis}}(T) $
to represent the SMS\ osmotic pressure below $T_{\text{c}}.$ One can also
check that $Ts_{\text{dis}}^{(\text{S)}}\rightarrow0,$ as $T\rightarrow0.$\ 

We will only discuss the disordered polymer phase below for $0\leq$ $m<1$. It
is easily checked that the above stability conditions in (\ref{PolyStbl}) are
always satisfied for $\lambda_{\text{dis}};$ see, for example, the behavior of
$\phi_{\text{B}}$ in Fig. 2, where we have taken $m=0.7,$ and $J=1$. Since the
high-temperature disordered phase represents a physical system, it cannot give
rise to a negative entropy $s_{\text{dis}}^{(\text{P)}}$ above $T_{\text{c}}$;
however, its metastable extension violates it as shown in Fig. 2, where its
entropy $s_{\text{dis}}^{(\text{P)}}$ becomes negative below $T_{\text{K}%
}\cong0.266,$ which is lower than the transition temperature $T_{\text{c}}.$

We now make an important observation. As $m$ decreases (below 1), both
$T_{\text{K}}$ and $T_{\text{c}}$ ($T_{\text{K}}<T_{\text{c}})$ move down
towards zero simultaneously. As $m\rightarrow0,$ the equilibrium ordered phase
corresponding to $\lambda_{\text{ord}}$\ disappear completely, and the
disordered phase corresponding to $\lambda_{\text{dis}}$ becomes the
equilibrium phase. There is no transition to any other state. Thus, there is
no metastability anymore. Consequently, there is \emph{no} Kauzmann point
since there is \emph{no} other ordered state any more, as argued above. Thus,
our exact calculation confirms our earlier conclusion that the existence of an
ordered state is crucial for the existence of the entropy crisis. The
existence of an ordered state sets the zero of the temperature scale by its
minimum energy $E_{0}$. This scale then sets the temperature $T_{\text{K}}$ of
the lowest SMS energy $E_{\text{K}}>E_{0}$ to be positive. \ Thus, one must
consider the ordered and the metastable states together.

We also observe that there is no singularity in $\lambda_{\text{dis}}$ or
$\omega_{\text{dis}}(T)$ at $T_{\text{c}}$, even though there is a phase
transition there.\ Similarly, there is no singularity in $\lambda_{\text{ord}%
}$ or $\omega_{\text{ord}}(T)$ at $T_{\text{c}}.$\ Thus, the thermodynamic
singularity in the equilibrium free energy does not necessarily create a
singularity in $\omega_{\text{dis}}(T)$ or $\omega_{\text{ord}}(T)$ at
$T_{\text{c}},$ as was discussed earlier. The existence of a singularity or
spinodal at some other temperature is a different matter.

It should be noted that the eigenvalues $\lambda_{\text{dis}}$ and
$\lambda_{\text{ord}}$ are independent of the size of the lattice. Therefore,
they can be used to describe not only the disordered and ordered phases, but
also the SMS, which is the continuation of the disordered phase, even for a
finite $N$. Thermodynamic limit is not necessary. For finite $N$, $Z_{\alpha
}(K,m)<Z(K,m)$ and the inequality becomes an equality in the sense of
(\ref{HF00}) only as $N\rightarrow\infty$ for the proper choice of $\alpha$
depending on the temperature.

\subsection{Binary Mixture Model}

We now consider\ a simple lattice model of an incompressible binary mixture of
two kinds of particles A and B, to be represented by an Ising spin $S$. The
two spin states (+1 or up) and ($-$1 or down) represent the particles of two
species A and B, respectively. As we are not interested in their phase
separation, but in the possibility of a glass transition, we assume that their
mutual interaction is attractive. In addition, we are interested in a
first-order transition for conventional supercooling. We will, therefore, use
an anti-ferromagnetic Ising model in zero magnetic field with both two-spin
($J>0$), and three-spin interactions ($J^{\prime}\neq0$); the latter ensures
that the melting transition is first order. In order to solve the model
exactly, we consider a Husimi cactus made of squares, on which the model can
be solved exactly\cite{GujPRL}. We consider the simplest cactus in which only
two squares meet at a site; they cannot share a lattice bond. The squares are
connected so that there are no closed loops except those formed by the
squares. The cactus can be thought as an approximation of a square lattice, so
that the exact Husimi cactus solution can be thought of as an approximate
solution of the square lattice model. There is a sublattice structure at low
temperatures caused by the anti-ferromagnetic interaction:\ particles of one
species are found on one of the two sublattices. We identify this ordered
structure as a crystal. The interaction energy is%

\begin{equation}
E\mathcal{=}J\sum SS^{\prime}+J^{\prime}\sum SS^{\prime}S^{\prime\prime}.
\label{BinaryE}%
\end{equation}
The first sum is over nearest-neighbor spin pairs and the second over
neighboring spin triplets within each square. In the absence of the three-spin
coupling, the two-spin coupling gives rise to an antiferromagnetic (AF)
ordering at low temperatures. For $J^{\prime}>(-J),$ the AF ordering remains
the preferred ordering, while for $J^{\prime}<(-J),$\ the ferromagnetic
ordering is prefereed. Threrefore, we only consider $J^{\prime}>(-J)$ in the
following. We set $J$=1 to set the temperature scale.

The model is solved recursively, as has been described elsewhere\cite{GujPRL}.
We label sites on the cactus by an index $m$, which increases\ sequentially
outwards from $m=0$ at the origin. We introduce partial PF's $Z_{m}(\uparrow)$
and $Z_{m}(\downarrow),$ depending on the states of the spin at the $m$-th
cactus level. It represents the contribution of the part of the cactus above
that level to the total PF. We then introduce the ratio
\begin{equation}
x_{m}\equiv Z_{m}(\uparrow)/[Z_{m}(\uparrow)+Z_{m}(\downarrow)],
\label{Ratios}%
\end{equation}
which satisfies the recursion relation
\begin{equation}
x_{m}\equiv f(x_{m+1},x_{m+2},v)/[f(x_{m+1},x_{m+2},v)+f(y_{m+1}%
,y_{m+2},1/v)], \label{RR}%
\end{equation}
where
\begin{equation}
f(x,x^{\prime},v)\equiv x^{2}x^{\prime}/u^{4}v^{4}+2xx^{\prime}yv^{2}%
+x^{2}yv^{2}+u^{4}x^{\prime}y^{2}+2xyy^{\prime}+y^{2}y^{\prime}/v^{2},
\label{function}%
\end{equation}
with
\[
u\equiv e^{\beta},v\equiv e^{\beta J^{\prime}},y\equiv1-x,y^{\prime}%
\equiv1-x^{\prime}.
\]

There are two kinds of fix-point solutions of the recursion relation that
describe the bulk behavior\cite{GujPRL}. In the 1-cycle solution, the
fix-point solution becomes independent of the level index $m$ as we move
towards the origin $m=0$, and is represented by $x^{\ast}.$ For the current
problem, it is given by $x^{\ast}=1/2$, as can be checked explicitly by the
above recursion relation in (\ref{RR}). It is obvious that it exists at all
temperatures. There is no singularity in this fix-point solution. This
solution corresponds to the disordered paramagnetic phase at high temperatures
and the SMS below the melting transition. The other fix-point solution is a
2-cycle solution, which has been found and discussed earlier in the
semi-flexible polymer problem$\cite{GujCorsi,GujRC}$, the dimer
model$\cite{FedorDimer},$ and star and dendrimer solutions$\cite{CorsiThesis}%
.$ The fix-point solution alternates between two values $x_{1}^{\ast},$ and
$x_{2}^{\ast}$ on two successive levels. At $T=0,$ this solution is given
either by $x_{1}^{\ast}=1,$ and $x_{2}^{\ast}=0,$ or by $x_{1}^{\ast}=0,$ and
$x_{2}^{\ast}=1.$\ The system picks one of these as the solution. At and near
$T=0,$ this solution corresponds to the low temperature AF-ordered phase,
which represents the CR and its excitation at equal occupation, and can be
obtained numerically. The 1-cycle free energy is calculated by the general
method proposed in \cite{GujPRL}, and the 2-cycle free energy is calculated by
the method given in \cite{GujCorsi,CorsiThesis}.

For $J^{\prime}=0.01,$ we find that $T_{\text{M}}\cong2.753$, where there is a
discontinuity in the entropy per site of 0.0168. The SMS below $T_{\text{M}}$
represents SCL, whose entropy density, see Fig. 2, vanishes at $T_{\text{K}%
}\cong1.132,$ and whose specific heat (not shown) remains positive everywhere
with a maximum at $T\cong1.26$. At absolute zero, the entropy per site
$S_{\text{dis}}\simeq-0.3466,$ while the CR entropy is zero, as expected$.$
The CR and SCL free energies per site become identical ($=2J$) at absolute
zero in accordance with the Theorem 3. Thus, the free energy diagram we obtain
in this case is similar to that in the inset in Fig. 1.

\section{Discussion \& Conclusions}

\subsection{Thermodynamic Criterion for Ideal Glass Transition}

The work was motivated by a desire to identify a general thermodynamic
mechanism for the glass transition. For this, we identify a general
thermodynamic condition. This general principle has also been verified in some
recent work on lattice models that have been carried
out\cite{FedorDimer,CorsiThesis}, and has its foundation in the original idea
of the entropy crisis $S_{\text{ex}}<0$ noted by Kauzmann, and is as follows.
The entropy due to a set of coupled degrees of freedom, when properly defined
either using quantum mechanics or by discretization as in a lattice model, can
never be negative, since the number of configurations for a given set of
macroscopic quantities must be an integer $\geq1.$ Consequently, this
condition of non-negative entropy $S$ is nothing but the condition of reality.
Its violation gives rise to the concept of absolute entropy crisis. It is in
this sense we have used the entropy crisis in this work. Such a notion is more
stringent than the Kauzmann criterion that $S_{\text{ex}}$ be not negative,
for which there is no thermodynamic justification, as we have argued in the
Introduction. For the glass transition, we are interested in the set of
coupled degrees of freedom that contains the configurational (i.e.,
positional) degrees of freedom. The entropy of such a set, regardless of
whether quantum or classical mechanics is used in its calculation, is what we
call the configurational entropy.

It should be noted that there are various other definitions of the
configurational entropy in the literature. Many workers take $S_{\text{ex}}$
to denote the configurational entropy. However, as said above, there is no
thermodynamic basis for $S_{\text{ex}}$ to be non-negative. So, it cannot play
any role in a thermodynamic theory of glass transition. The landscape picture
identifies the entropy $S_{\text{IS}}$ of the inherent structure with the
configurational entropy. Its vanishing is used to identify the glass
transition. It is easy to see\cite{GujFedor} that for the classical
configurational PF, the two entropies are related:
\[
S(T)=S_{\text{IS}}(T)+S_{\text{basin}}(T),
\]
where $S_{\text{basin}}(T)\equiv\partial(T\ln\overline{Z}_{\text{basin}%
})/\partial T$ is the entropy arising from the average basin PF $\overline
{Z}_{\text{basin}}(T)$ in the landscape picture. Thus, our criterion $S(T)=0$
of the entropy crisis is also stringent than $S_{\text{IS}}(T)=0$ in the
landscape picture. Our criterion will also require $S_{\text{basin}}(T)=0,$
which can only occur at a temperature lower than the temperature at which
$S_{\text{IS}}(T)=0.$ It should be noted that there is no kinetic energy
contribution in $S_{\text{basin}}(T)$ as the landscape picture deals only with
the classical configurational PF; the translational degrees of freedom are
decoupled in classical mechanics, as discussed above.

\subsection{Continuum vs. Discrete Models}

The best known example of classical models giving rise to negative entropy at
low temperatures is the ideal gas. Similarly, classical real gases also give
rise to negative entropy at low temperatures. This problem can be easily
traced to the fact that we are treating the real and momentum spaces as
continuum\cite{FedorDimer}. Another well-known example is the Tonks gas of
rods in one-dimension (no kinetic energy), which also gives negative entropy
at high coverage\cite{Tonks,FedorDimer}. Here, the one-dimensional space is
treated as continuum. On the other hand, it is well known that a classical
lattice model will never give rise to negative entropy. Similarly, the random
energy model\cite{Derrida}, which treats energy as continuum, gives negative
entropy at low temperatures.

It is clear that the problem of negative entropy is not due to the classical
nature of the systems, but due to the continuum nature of the
model\cite{Note2}. To ensure non-negative entropy, we must discretize the
model, as we have discussed here. Once this has been done, the entropy crisis
becomes a genuine crisis imposed by the reality condition as we have proposed earlier

\subsection{General Thermodynamic Mechanism for Ideal Glass Transition}

The general thermodynamic mechanism of the ideal glass transition occurs in
any system that has an ordered state, distinct from the high-temperature
disordered state. The continuation of the free energy of the disordered state
below the melting transition at $T_{\text{M}}$ gives the free energy of the
stationary metastable state. We have shown that this continuation always gives
rise to a stable free energy. For example, it gives rise to a non-negative
heat capacity for $T\geq0$. This SMS free energy $f_{\text{dis}}(T)$ becomes
equal to $E_{0}$ at $T_{\text{eq}}>0$, as the temperature is reduced. The
energy continues to drop until finally, $f_{\text{dis}}(T)=E_{0} $ at
$T=0.$\ Since $f_{\text{dis}}(T)=E_{0}$ at $T=0$ and at $T=T_{\text{eq}} $, it
must have a maximum at some intermediate temperature $0<T_{\text{K}%
}<T_{\text{eq}}.$ The energy $E_{\text{K}}>E_{0}$ at $T_{\text{K}}$ because of
positive heat capacity. The entropy is zero at $T_{\text{K}}$ and negative
below $T_{\text{K}}$. Thus, the SMS over the range $(0,T_{\text{K}}) $ is
unphysical and must be replaced by the ideal glass.

The above mechanism has been shown to be generic by the rigorous analysis,
which is valid for classical and quantum systems. To the best of our
knowledge, this result is the first of its kind and shows that the entropy
crisis is genuine in those systems in which there is a more stable ordered
state than the disordered SMS.

We have also considered models\cite{FedorDimer} in which there are two
possible transitions; one of them is from a disordered phase to an
intermediate phase, and the second one at a lower temperature from the
intermediate phase to an ordered phase. In this case, two possible SMS's as
the continuation of the disordered and intermediate phases emerge, with. each
extension giving rise to its own entropy crisis; in addition, both have the
same free energy as $T=0$ as the ordered phase. Thus, the mechanism is generic.

\subsection{SMS \& Exact Calculations}

The transition between SMS and the ideal glass is not brought about by any
thermodynamic singularity at $T_{\text{K}};$ rather, it is \emph{imposed} by
the reality requirement. To the best of our knowledge, the ideal glass state
does \emph{not} explicitly emerge as a phase in any calculation that has been
carried out so far, including the two that are presented here. In this sense,
this transition is a very special kind of transition, which does not seem to
belong to the class of phase transitions in which various phases emerge in the calculation.

The two examples that we have presented here show the existence of SMS. Thus,
they demonstrate that our hypothesis of SMS existence is not vacuous. Both
examples also show genuine entropy crisis in SMS. Thus, they provide support
for the violation of the reality condition in SMS at a positive temperature in
exact calculations. We need to invoke an ideal glass transition at this
temperature in each model.

The one-dimensional exact calculation is not a mean-field type calculation,
and is presented not only to overcome the folklore that SMS's exist only at
the mean-field level, but also to explicitly confirm the theorems. It is a
model of a polymer system and confirms all the theorems. In particular, it
shows that as $m\rightarrow0$, the ideal glass transition disappears because
there is no ordered state anymore. Hence, our results are not of mean-field nature.

The second example is also presented not only to overcome the folklore that
frustration is crucial in the transition, but also to establish that entropy
crisis is possible in systems containing small molecules, and not just
polymers. There is no frustration in the Ising model because the cactus
consists of squares. This example also confirms the theorems and the corollary.

Both examples show that the ideal glass transition occurs at a positive
temperature and the energy $E_{\text{K}}$ at that temperature is higher than
$E_{0},$ the ideal CR\ energy at $T=0$. This means that the ideal glass has a
higher energy than the crystal at absolute zero, in conformity with the experiments.

\subsection{Absence of Entropy Crisis in Equilibrium State}

We now come to a very important consequence of our corollary. It is the
following. The presence of an entropy crisis at a positive temperature implies
that there must exist an equilibrium state for which no entropy crisis can
exist. The equilibrium states in any system or model calculation, if carried
out exactly, will \emph{never} give rise to any entropy crisis at a positive
temperature. This is because the lowest energy $E_{0}$ determines the lowest
allowed temperature $T=0$ in the system, even for a finite system. Since there
cannot be any singularity in any finite-system equilibrium free energy, the
latter should continue all the way down to $T=0$ and the equilibrium entropy
must remain non-negative at all temperatures; the latter can only vanish at
$T=0.$ Thus, there would be no entropy crisis at a positive temperature in the
equilibrium state. This will remain true as the thermodynamic limit is taken.
Thus, \emph{no} entropy crisis can occur at a finite temperature in the
equilibrium state in that the entropy becomes \emph{negative} below that
temperature. However, it is possible that as $N\rightarrow\infty,$ the
equilibrium free energy becomes horizontal, so that the entropy vanishes, over
a non-zero temperature range $(0,T_{\text{C}}).$ Since the equilibrium free
energy exists with non-negative entropy for all $T\geq0$ for finite $N$, the
free energy must show a singularity at $T_{\text{C}}.$ While the system is
frozen over the range $(0,T_{\text{C}}),$ its appearance is accompanied by a
phase transition. (This should be contrasted with the existence of a Kauzmann
point, below which the entropy becomes negative, but its appearance is not
accompanied by any singularity in the SMS free energy. Replacing the
unphysical SMS free energy below the Kauzmann point by a frozen state is done
by hand; it does not emerge as part of the calculation.) Thus,we conclude that
equilibrium state in any system will never show an entropy crisis at a
positive temperature. The zero of the temperature scale is determined by the
lowest possible energy $E_{0}.$ If any exact calculation for the free energy
or the entropy predicts an entropy crisis at a positive temperature, this will
necessarily imply that there must exist another state, the equilibrium state,
which will not show an entropy crisis.

This observation has been crucial in a recent investigation of a dimer model
\cite{FedorDimer} in which the disordered phase underwent a first-order
transition to an equilibrium ordered phase. The ordered phase then gave rise
to an entropy crisis at a lower temperature, which forced us to look for
another equilibrium state, which was eventually discovered above the
temperature where the entropy crisis was found, so that the crisis occurred in
a metastable state (this time emerging form an intermediate ordered state) as
we have suggested and followed the mechanism proposed here.

\subsection{Landscape Picture}

Finally, we wish to make connection of the ideal glass energy $E_{\text{K}}$
with the inherent structure in the landscape picture. The ideal glass at $T=0
$ in the canonical ensemble must be at a local minimum in the landscape.
Hence, its energy must be the energy of the particular inherent structure.
Since the ideal glass emerges at $T_{\text{K}}$, where $S(T)=0,$ we are forced
to conclude that this inherent structure also represents SMS at $T_{\text{K}%
}.$ This most certainly implies that SMS must be confined in the basin
associated with the SMS inherent structure. This confinement must occur at a
higher temperature (than $T_{\text{K}})$ where $S_{\text{IS}}(T)=0.$

In conclusion, we have justified the mechanism that gives rise to an entropy
crisis in metastable states in systems in which there exists is a more stable
phase. The generality of the mechanism is reflected in the generality of the
validation of the mechanism, which is common in classical and quantum
mechanical systems.

We would like to thank Andrea Corsi and Fedor Semerianov for various useful
discussions, and help with the figures (Andrea Corsi).

\begin{center}
{\large Figure Captions}
\end{center}

\begin{enumerate}
\item  Schematic form of the generic entropy functions for various possible states.

\item  The bond and the entropy densities. Both models show an entropy crisis
at a positive temperature.
\end{enumerate}

\bigskip
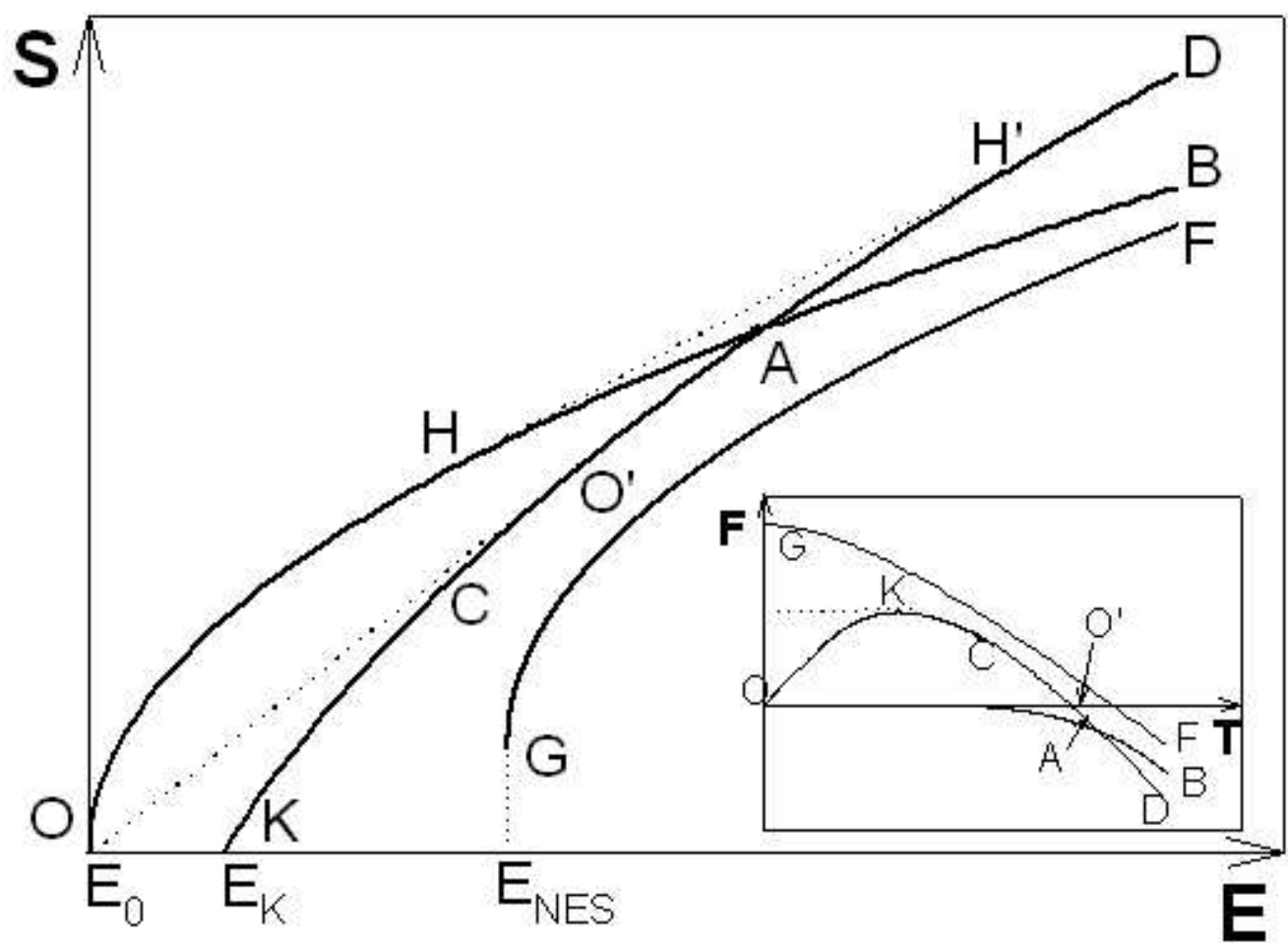
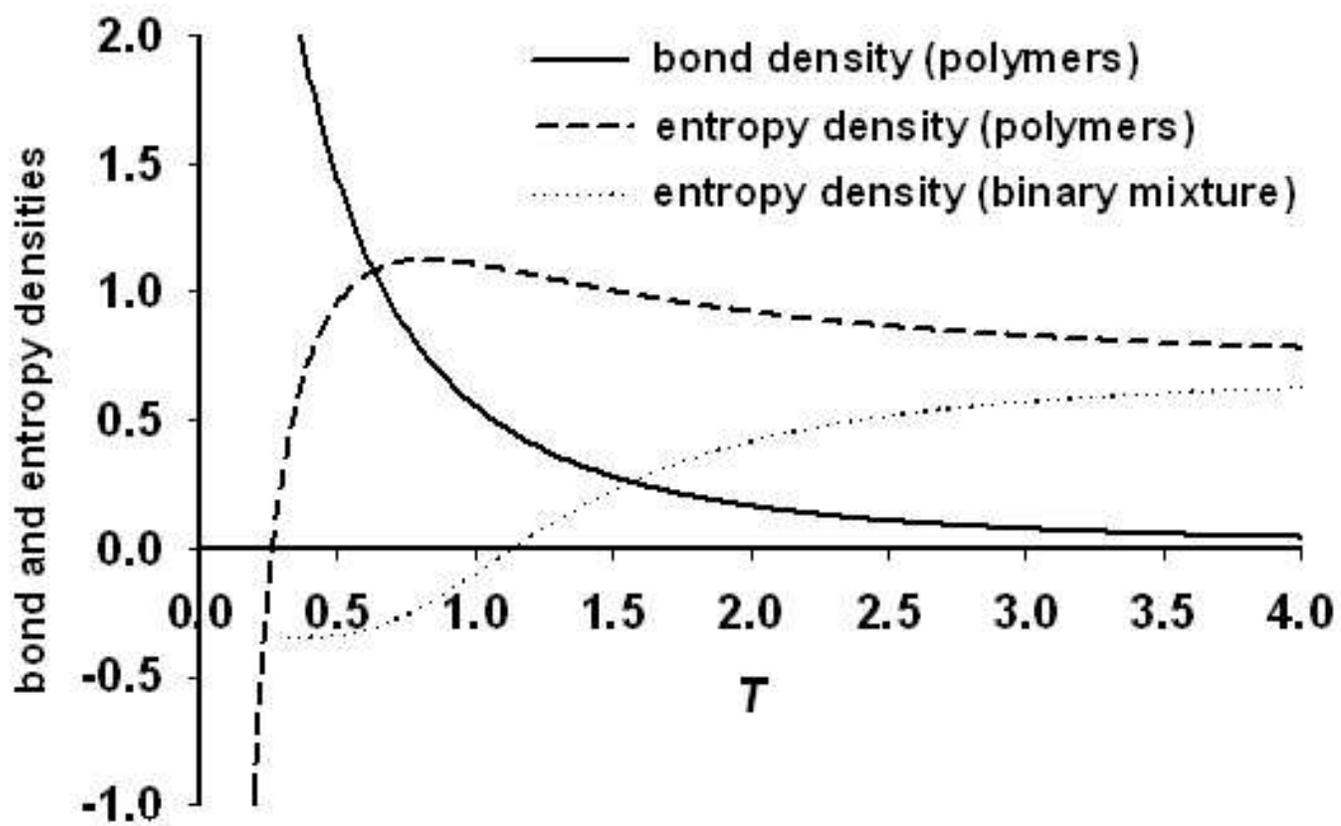
\end{document}